\documentclass[useAMS,usenatbib]{mn2e}

\usepackage{natbib}
\usepackage{latexsym,graphicx}
\usepackage{color}
\usepackage{fixltx2e}
\usepackage{verbatim} \usepackage{float}
\usepackage{amsmath,amssymb}
\usepackage{times}
\usepackage{setspace}
\usepackage{rotating}
\usepackage{comment}
\usepackage{pdflscape}
\usepackage{subfig}
\usepackage[squaren, Gray, cdot]{SIunits}

\usepackage{hyperref}
\def\apgt{\ {\raise-.5ex\hbox{$\buildrel>\over\sim$}}\ }
\def\aplt{\ {\raise-.5ex\hbox{$\buildrel<\over\sim$}}\ }

\let\oldhat\hat
\renewcommand{\hat}[1]{\oldhat{\mathbf{#1}}}

\renewcommand{\degree}{\ensuremath{^\circ}}

\title[Bipolarity in Wolf-Rayet nebulae]{On the bipolarity of Wolf-Rayet nebulae}

\author[D. M.-A.~Meyer et al.]
       {D. M.-A.~Meyer\thanks{E-mail: dmameyer.astro@gmail.com} \\
       Universit\" at Potsdam, Institut f\" ur Physik und Astronomie, 
       Karl-Liebknecht-Strasse 24/25, 14476 Potsdam, Germany\\
       }

\begin{document}

\date{Received; accepted}

\maketitle

\label{firstpage}

\begin{abstract} 
Wolf-Rayet stars are amongst the rarest but also most intriguing massive stars. 
Their extreme stellar wind\textcolor{black}{s} induce famous multi-wavelength circumstellar 
gas nebulae of various morphologies, spanning from circles and rings to bipolar 
shapes. 
This study is devoted to the investigation of the formation of young, asymmetric 
Wolf-Rayet gas nebulae and we present a 2.5-dimensional magneto-hydrodynamical 
toy model for the simulation of Wolf-Rayet gas nebulae generated by wind-wind 
interaction. 
Our method accounts for stellar wind asymmetries, rotation, magnetisation, evolution 
and mixing of materials. 
It is found that the morphology of the Wolf-Rayet nebulae of blue supergiant 
ancestors is tightly related to the wind geometry and to the stellar phase 
transition time interval, generating either a broadened peanut-like or a 
collimated jet-like gas nebula. 
Radiative transfer calculations of our Wolf-Rayet nebulae for dust infrared 
emission at $24\, \mu \rm m$ show that the projected diffuse emission can appear 
as oblate, bipolar, ellipsoidal or ring structures. Important projection 
effects are at work in shaping observed Wolf-Rayet nebulae. \textcolor{black}{This} 
might call a revision of the various classifications of Wolf-Rayet shells, 
\textcolor{black}{which are mostly based} on their observed shape. 
Particularly, our models question the possibility of producing pre-Wolf-Rayet wind 
asymmetries, responsible for bipolar nebula\textcolor{black}{e} like NGC 6888, within 
the single red supergiant evolution channel scenario. 
We propose that bipolar Wolf-Rayet nebulae can only be formed within the red 
supergiant scenario by multiple/merged massive stellar systems, or by 
single high-mass stars undergoing additional, e.g. blue supergiant, 
evolutionary stages prior to the Wolf-Rayet phase. 
\end{abstract}



\section{Introduction}
\label{sect:intro}

Formed out of the gravitational collapse of opaque pre-stellar 
cores, high-mass stellar objects, i.e. 
with mass $\ge8\, \rm M_{\odot}$, are cosmic regulators 
of the cycle of matter at work in the interstellar medium (ISM) 
of galaxies~\citep{maeder_2009,langer_araa_50_2012}. 
One of their characteristic feature\textcolor{black}{s} is the fast and strong wind  
blown out from their surface and powered by the complex nuclear 
reactions at work in their interior. 
Throughout their lives, massive stars first experience a rather 
long, main-sequence phase corresponding to the burning of hydrogen in 
their cores~\citep{ekstroem_aa_537_2012}. 
It is characterised by the release of diluted, hot and supersonic winds 
from the stellar surface, and, once their hydrogen is exhausted, stellar 
evolution is triggered. 
The entire evolution of massive stars is fixed by their zero-age 
main-sequence properties, such as their mass, angular velocity and chemical 
composition, which uniquely determines their evolution and ultimate 
fate~\citep{woosley_rvmp_74_2002,vink_asp_353_2006,
brott_aa_530_2011a,2020arXiv200408203S}. 
These initial stellar characteristics rule the time-dependence of 
the stellar surface properties, \textcolor{black}{as well as the number and} duration of 
the various evolutionary phases, up to their death, either as a core-collapse 
supernova leaving behind a plerionic supernova 
remnant, or directly collapsing as a black hole~\citep{Chevalier_araa_15_1977, 
weiler_araa_25_1988,woosley_rvmp_74_2002,woosley_araa_44_2006,smartt_araa_47_2009,
mueller_mnras_448_2015,gabler_mnras_502_2021}.

Amongst the many possible evolutionary histories massive stars can 
undergo, high-mass stellar objects can evolve to the so-called Wolf-Rayet 
stage. This phase is characterised 
by fast ($\sim 1000$-$5000\, \rm km\, \rm s^{-1}$) and dense stellar 
winds expelled at mass-loss rates $\ge 10^{-5}\, \rm M_{\odot}\, \rm yr^{-1}$ 
and enriched in C, N and O elements~\citep{Hamann2006,Sander2012,bestenlehner_aa_570_2014}. 
The precise manner a massive stellar object acquires the spectroscopic 
properties of a Wolf-Rayet star is not completely understood, 
however, a few typical evolutionary paths have been identified~\citep{crowther_araa_45_2007}. 
The most common channel reposes in an evolution from the main-sequence to the Wolf-Rayet 
phase through a red supergiant phase, and it mostly concerns Galactic stars with masses 
$\ge 20\, \rm M_{\odot}$. 
Heavier zero-age main-sequence stars (up to $\sim 60\, \rm M_{\odot}$) have been 
shown theoretically to experience a much more complex evolution, including 
several successive luminous blue variable phases, during which the star 
blows, sometimes eruptively, blue supergiant material~\citep{groh_aa564_2014}.

The interaction of the stellar wind of massive stars with the ISM 
produces gas nebulae witnessing the release of energy, momentum and 
heavy chemical elements which enriched their local ambient 
medium~\citep{langer_araa_50_2012}. 
The classical picture for stellar wind-ISM interaction is that 
of~\citet{weaver_apj_218_1977} and it consists of a several parsec-scale 
structured circumstellar bubbles surrounding massive stars. 
The circumstellar medium of Wolf-Rayet stars is different, as many of 
them display unusual smaller-scale ring-like gas nebulae detected in 
H$\alpha$ and infrared, both in the Milky Way, see~\citet{chu_apj_250_1981} 
and~\citet{treffers_apj_254_1982}, and in the Magellanic 
Clouds~\citep{chu_apj_255_1982,chu_apj_269_1983,Dopita_ApJS_93_1994,
weis_aa_325_1997,hung_apjs_252_2021}, respectively. A classification 
of circumstellar structures around Wolf-Rayet stars arose from that 
studies~\citep{chu_apj_249_1981,chu_apj_255_1982,chu_apj_254_1982,
treffers_apj_254_1982,chu_apjs_53_1983,toala_aa_578_2015}. 
Several follow-up surveys deepened and completed these observations \textcolor{black}{in the} 
southern hemisphere, see~\citet{marston_apjs_93_1994} and~\citet{1994_apj_marston_95}. 
Wolf-Rayet nebulae entered the X-rays domain with 
{\it Asca}~\citep{wrigge_apj_633_2005}
{\it Rosat} and  
{\it Chandra}~\citep{guerrero_apjs_177_2008},  
{\it XMM-Newton}~\citep{toala_apj_755_2012} 
observations from their (shocked) stellar winds witnessing 
emission of optically-thin high temperature plasma.  
More recently, a radio non-thermal synchrotron counterpart of the stellar 
wind bubble of a Wolf-Rayet ring nebula indicated that these circumstellar 
nebulae are the site of particles accelelration~\citep{prajapati_apj_884_2019}. 
This accumulation of observations motivates the present numerical efforts 
aiming at understanding the peculiar vicinity of young Wolf-Rayet stars.

Numerical simulations rapidly became an efficient tool to understand the 
different physical processes at work in the circumstellar medium of massive 
stars~\citep{comeron_aa_338_1998,freyer_apj_594_2003,dwarkadas_apj_630_2005,
freyer_apj_638_2006,toala_apj_737_2011,vanmarle_apj_734_2011,vanmarle_584_aa_2015}, 
to constrain stellar evolution models/probe local ISM 
conditions~\citep{mackey_apjlett_751_2012,2014Natur.512..282M}, 
and to understand the pre-supernova surroundings of dying massive 
stars, with which supernovae shock wave\textcolor{black}{s} subsequently 
interact~\citep{ciotti_aa_215_1989,chevalier_apj_344_1989, 
vanmarle_aa_444_2005,vanmarle_aa_460_2006,meyer_mnras_502_2021}. 
Particularly, the gas nebula\textcolor{black}{e} formed around Wolf-Rayet stars have been 
studied numerically as evolved winds, colliding with the material of a 
larger-scale main-sequence wind-blown bubble~\citep{brighenti_mnras_285_1997, 
dwarkadas_apj_667_2007,vanmarle_aa_469_2007,vanmarle_bull_srs_liege_80_2011}.

It has recently been shown in~\citet{meyer_mnras_496_2020}, on the basis of 
magneto-hydrodynamical (MHD) simulations, that (i) the circumstellar rings appearing 
\textcolor{black}{to} be comoving with the fastest-moving Wolf-Rayet stars are a trace of their very high 
initial masses ($\ge35$ to $\sim 60\, \rm M_{\odot}$) inducing complex evolutionary 
histories, and that (ii) their large-scale surroundings are main-sequence stellar 
wind bow shock\textcolor{black}{s} fainter than the inner rings which became unobservable as the driving 
star runs \textcolor{black}{into} diluted media~\citep{2010MNRAS.405.1047G}. 
This solved the apparent missing bow shock problem around some high-latitude, 
Galactic, very fast-moving Wolf-Rayet stars like WR124, surrounded by its compact 
gas nebula M1-67~\citep{chu_apj_249_1981,sluys_aa_398_2003,
marchenko_apj_724_2010,Toala2018} or WR71~\citep{Faherty_apj_147_2014,Flagey_aj_148_2014}. 
If we can now qualitatively explain the formation of quasi-spherical rings 
around very fast-moving Wolf-Rayet stars, detailed observations of these 
gas nebulae reveal much more complex structures \textcolor{black}{whose formation 
are far} from being understood. 
Which mechanisms distinguish between the production of \textcolor{black}{ring-like} shells 
from ovoidal and/or other complex shapes ? What can we learn \textcolor{black}{regarding 
the} past evolution of massive stars with nebulae of such morphologies ?

In this study, we continue our investigations of the shaping of circumstellar 
Wolf-Rayet nebulae started in~\citet{meyer_mnras_496_2020}. We concentrate on 
the production of bipolar asymmetries and non-spherical morphologies observed 
in the vicinity of some young Wolf-Rayet stars~\citep{toala_aa_578_2015}. 
The starting point of this study is the work of~\citet{brighenti_mnras_285_1997},  
in which the development of young aspherical Wolf-Rayet nebulae is examined using an 
hydrodynamical wind-wind interaction model. Their simulations focus on the particular 
scenario of a red supergiant star, with asymmetric stellar wind properties, evolving 
to a Wolf-Rayet stage of symmetric stellar wind properties. 
In the same spirit, we build a magneto-hydrodynamical toy model for rotating 
blue supergiant stars with asymmetric stellar winds and evolving to the Wolf-Rayet 
phase. The effects of asymmetries in the blue supergiant wind are investigated 
using the recipe of~\citet{raga_apj_680_2008}. We perform near-infrared synthetic 
observables of the corresponding bipolar stellar wind nebulae, in order to 
discuss their respective emission properties and facilitate qualitative 
comparison to observations available in the literature.

Our study is organised as follows. Firstly, we present the numerical
methods utilised for the MHD simulations of Wolf-Rayet gas nebulae 
evolving after having undergone a previous blue supergiant 
phase in Section~\ref{sect:method}. 
We show our results regarding to the dynamical evolution of bipolar MHD 
Wolf-Rayet nebulae generated by wind-wind interaction in Section~\ref{sect:results}. 
The results are further discussed in Section~\ref{sect:discussion}, 
and finally, we present our conclusions in Section~\ref{sect:conclusion}.


\section{Method}
\label{sect:method}

This section presents the numerical methods used in this project, together with 
the initial conditions and parametrised boundary conditions of the simulations. 
Last, we introduce the reader to the models performed in this study.

\subsection{Governing equations}
\label{sect:method_equation}

We set our problem in the frame of non-ideal magneto-hydrodynamics, described 
by the following series of equations, 
\begin{equation}
	   \frac{\partial \rho}{\partial t}  + 
	   \bmath{\nabla}  \cdot \big(\rho\bmath{v}) =   0,
\label{eq:mhdeq_1}
\end{equation}
\begin{equation}
	   \frac{\partial \bmath{m} }{\partial t}  + 
           \bmath{\nabla} \cdot \Big( \bmath{m} \textcolor{black}{\otimes} \bmath{v}  
           \textcolor{black}{-} \bmath{B} \textcolor{black}{\otimes} \bmath{B} + \bmath{\hat I}p_{\rm t} \Big)  
            =   \bmath{0},
\label{eq:mhdeq_2}
\end{equation}
\begin{equation}
	  \frac{\partial E }{\partial t}   + 
	  \bmath{\nabla} \cdot \Big( (E+p_{\rm t})\bmath{v}-\bmath{B}(\bmath{v}\cdot\bmath{B}) \Big)  
	  = \Phi(T,\rho),
\label{eq:mhdeq_3}
\end{equation}
and,
\begin{equation}
	  \frac{\partial \bmath{B} }{\partial t}   + 
	  \bmath{\nabla} \cdot \Big( \bmath{v}  \textcolor{black}{\otimes} \bmath{B} - \bmath{B} \textcolor{black}{\otimes} \bmath{v} \Big)  =
	  \bmath{0},
\label{eq:mhdeq_4}
\end{equation}
where $\bmath{B}$ represents the magnetic field vector, $\rho$ stands for the mass 
density, 
\begin{equation}
    \bmath{m}=\rho\bmath{v}, 
\end{equation}
is the linear momentum vector, $\bmath{\hat I}$ the identity matrix, 
$p_{\rm t}$ is the total pressure and $v$ the gas velocity, respectively. 
The total energy of the system reads, 
\begin{equation}
	E = \frac{p}{(\gamma - 1)} + \frac{ \bmath{m} \cdot \bmath{m} }{2\rho} 
	    + \frac{ \bmath{B} \cdot \bmath{B} }{2},
\label{eq:energy}
\end{equation}
and the sound speed is defined as, 
\begin{equation}
	  c_{\rm s} = \sqrt{ \frac{\gamma p}{\rho} },
\label{eq:cs}
\end{equation}
which closes the above system of equations, and where $\gamma=5/3$ 
is the adiabatic index.

Radiative cooling and heating of the gas,  
\begin{equation}  
	 \itl \Phi(T,\rho)  =  n_{\mathrm{H}}\Gamma(T)   
		   		 -  n^{2}_{\mathrm{H}}\Lambda(T),
\label{eq:dissipation}
\end{equation}
are explicitly included via optically-thin processes for loss 
$\Lambda(T)$ and gain $\Gamma(T)$ in the the source term, where,  
\begin{equation}
	T =  \mu \frac{ m_{\mathrm{H}} }{ k_{\rm{B}} } \frac{p}{\rho},
\label{eq:temperature}
\end{equation}
is the gas temperature, and using the laws of photoionized gas 
described in great detail in~\citet{2021arXiv210705513M}.  
Hence, our method ignores any explicit treatment of the radiation pressure 
from the photon field of the central star. 
\textcolor{black}{
This is correct} for cool blue supergiant stars, while we neglect the 
detailed position of the ionization front generated by the young 
Wolf-Rayet stars, as well as its effects on the shell 
instabilities~\citep{toala_apj_737_2011}.

A passive scalar tracer is \textcolor{black}{included in} the stellar wind, 
\begin{equation}
	\frac{\partial (\rho Q) }{\partial t } +  
	\bmath{ \nabla } \cdot  ( \bmath{v} \rho Q) = 0,
\label{eq:tracers}
\end{equation}
to follow the advection of the new-born Wolf-Rayet circumstellar nebula, 
as compared to the previous blue supergiant wind. 
This system of equations is solved by use of the eight-wave \textcolor{black}{algorithm~\citep{Powell1997},} 
within a second-order Runge-Kutta with parabolic reconstruction of the variables between 
neighbouring cells together with the HLL Riemann solver~\citep{hll_ref}. 
This unsplit scheme ensures that the magnetic field is 
divergence-free everywhere in the computational domain. The simulations 
timesteps are controlled by the \textcolor{black}{Courant-Friedrich-Levi} condition that we 
set to $C_{\rm cfl}=0.1$.

\begin{table}
	\centering
	\caption{
	Stellar surface parameters used in our two-winds models. 
	}
	\begin{tabular}{lccr}
	\hline
	$\mathrm{Quantities}$                  &
	$\mathrm{Symbol}$                      &	
	$\mathrm{Supergiant}$                  &
	$\mathrm{Wolf}$$-$$\mathrm{Rayet}$     \\
	\hline   
	Accretion rate        &  $\dot{M}\, (\rm M_{\odot}\, \rm yr^{-1})$      
	                      &  $10^{-6}$     & $10^{-4.3}$         \\ 
	Wind velocity         &  $v_{\rm w}\, (\rm km\, \rm s^{-1})$    
	                      &  500         & 1900    \\    
	Angular velocity      &  $v_{\rm rot}\, (\rm km\, \rm s^{-1})$  
	                      &  60          & 10    \\ 
	Surface field         &  $B_{\star}\, (\rm G)$    
	                      &  1           & 100    \\ 
	Stellar radius        &  $R_{\star}\, (\rm R_{\odot})$   
	                      &  20          & 2.3         \\ 	
	\hline    
	\end{tabular}
\label{tab:stars}
\\
\end{table}

\subsection{Initial conditions}
\label{sect:ic}

We carry out MHD simulations with the {\sc pluto} 
code~\citep{mignone_apj_170_2007,migmone_apjs_198_2012,vaidya_apj_865_2018}. 
The calculations are performed using a 2.5-dimensional spherical coordinate 
system ($O$;$r$,$\theta$,$\phi$) of origin $O$, with $r$ the radial direction, 
$\theta$ the poloidal coordinate and $\phi$ the toroidal component. 
The coordinate system $[r_{\rm in},r_{\rm out}]\times[0,\pi]\times[0,2\pi]$ 
is mapped with a mesh that is uniform along the polar and toroidal directions, 
while it expands logarithmically along the radial direction. 
It permits to reach high spatial resolutions close to the \textcolor{black}{stellar wind 
boundary $r_{\rm in}=0.02\, \rm pc$}, while reducing the total 
number of grid zones in the simulation domain and reducing the computational costs. 
In total, we use $200\times\,200\times\, 1$ grid zones. 
Outflow boundary conditions are assigned at the outer boundary 
\textcolor{black}{$r_{\rm out}=10\, \rm pc$}, reflective 
boundary conditions are imposed along the 
symmetry axis $\theta=0$ and $\theta=\pi$, while periodic boundary 
conditions are used at the borders $\phi=0$ and $\phi=2\pi$. 
\textcolor{black}{
Since the inner boundary is much larger than the radius of the photosphere 
($r_{\rm in} \gg R_{\star}$), stellar gravity and wind acceleration mechanisms  
such as coronal heating and radiation pressure, are neglected in the 
computational domain. Therefore, the acceleration of the stellar wind to the 
terminal velocity is not modelled in the simulations. 
}

The simulations are initialised with the asymmetric stellar wind of a blue 
supergiant star. The adopted latitude-dependence of the wind density is that 
developed in the context of the astrosphere of the runaway asymptotic giant 
branch star Mira~\citep{raga_apj_680_2008}. It reads, 
\begin{equation}
	\rho_{w}(r,\theta) = \frac{ \dot{M} }{ 4\pi r^{2} v_{\rm w} }
	\rightarrow \frac{ A }{ r^{2} }f( \theta ), 
\label{eq:wind}
\end{equation}
with $\dot{M}$ the stellar mass-loss rate, $v_{\rm w}$ the stellar wind 
velocity and $r$ the radial distance to the central star. The function, 
\begin{equation}
	f(\theta) = \xi - ( \xi - 1 ) | \cos( \theta ) |^{p}, 
\label{eq:f}
\end{equation}
measures the wind anisotropy, with $p$ a parameter determining the 
flattening degree of the density towards the equator and $\xi$ is 
the equator-to-pole density ratio, respectively. The scaling factor 
$A$ ensures that the latitude-dependence in the stellar wind conserves 
the amount of mass lost by the star per unit time. It is evaluated 
by writing the mass conservation at the stellar surface, 
\begin{equation}
    \dot{M } = 4\pi r^{2} \int_{0}^{\pi/2} \rho_{w}(r,\theta) v_{w}(r,\theta) \sin( \theta ) d\theta, 
    \label{eq:Mw1}
\end{equation}
yielding, 
\begin{equation} 
    \dot{M } = 4\pi A v_{\rm w} \int_{0}^{\pi/2} |f(\theta)|^{p} \sin( \theta ) d\theta,  
    \label{eq:Mw2}    
\end{equation}
where $v_{\rm w}$ is the stellar wind terminal velocity, and, finally, one 
obtains, 
\begin{equation}
    A = \frac{ \dot{M } }{ 4\pi v_{\rm w} } 
    \frac{ (\xi-1)^{2} }{ 8\xi^{5/2} - 20\xi + 12 }, 
    \label{eq:A}    
\end{equation}
using $p=1/2$. We chose this prescriptions for stellar wind asymmetries since it 
has already been used in the context of massive stars~\citep{fang_mnras_464_2017}, 
although other recipes exist~\citep{frank_apj_441_1995,blondin_apj_472_1996}.

\begin{table*}
	\centering
	\caption{
	List of models in our study. The table informs on the simulation 
	labels, whether the blue supergiant wind is symmetric or not, the blue supergiant 
	to Wolf-Rayet phase transition timescale $\Delta t$ (in $\rm yr$) and  
	the exponent $\beta$ controlling the variations of the different 
	quantities during the phase transition, the flattening degree of the 
	blue supergiant wind towards the equator $p$ and the equator-to-pole 
	density ratio $\xi$, respectively. The last column gives the 
	general purpose of each simulation models. 
	}
	\begin{tabular}{lcccccr}
	\hline
	${\rm {Model}}$ & $\mathrm{Asymmetric}\, \mathrm{wind}$  
	                & $\mathrm{\Delta t}\, (\rm yr)$   
	                &   $\beta$ 
	                &   $p$ 
	                &   $\xi$ 
	                &   $\mathrm{Purpose}$\\ 
	\hline   
	Run-Base                &  yes    & $0$      & $4$ & $1/4$ & $20$  & Baseline model \\ 
	Run-Sym                 &  no     & $0$      & $-$ & $-$   & $-$   & Model with isotropic progenitor wind \\ 
	Run-$\Delta $t10kyr    &  yes    & $10^{4}$  & $4$ & $1/4$ & $20$  & Model with $\Delta $t10kyr phase transition \\ 
	Run-$\xi$20            &  yes    & $10^{4}$  & $4$ & $1/4$ & $20$  & Model with phase transition and 
	lower pole-to-equator density contrast \\ 
	Run-$\xi$100           &  yes    & $10^{4}$  & $4$ & $3/4$ & $100$ & Model with phase transition and
	higher pole-to-equator density contrast   \\ 
	Run-$\beta$4           &  yes    & $5\times10^{4}$ & $4$   & $1/4$ & $20$  & Model with quadratic  
	interpolation between winds during the phase transition \\ 
	Run-$\beta$2           &  yes    & $5\times10^{4}$ & $4$   & $1/4$ & $20$  & Model with linear  
	interpolation between winds during the phase transition \\ 
	\hline    
	\end{tabular}
\label{tab:models}
\\
\end{table*}

Additionally,~\citet{raga_apj_680_2008} prescribe stellar wind flow 
assigned at the wind boundary $r_{\rm in}$ to account for the asymmetry 
function as,
\begin{equation}
	v(r_{\rm in},\theta) = \frac{ v_{\rm w} }{ \sqrt{ f(\theta) } },
\label{eq:vel}
\end{equation}
so that the ram pressure of the wind of the inner boundary, 
\begin{equation}
	\rho_{w}(r_{\rm in},\theta) v(r_{\rm in},\theta)^{2} 
	 = \frac{ A }{ r_{\rm in}^{2} }f( \theta ) 
	 \Big(   \frac{ v_{\rm w} }{ \sqrt{ f(\theta) } }  \Big)^{2} 
	 \propto \Big( \frac{ v_{\rm w} }{ r_{\rm in} } \Big)^{2},
\end{equation}
remains isotropic. 
Stellar rotation is included \textcolor{black}{in} the model by considering the 
following latitude-dependent equatorial rotation velocity,  
\begin{equation}
	v_{\phi}(\theta) = v_{\rm rot} \sin( \theta ),
\label{eq:Vphi}
\end{equation}
where $v_{\rm rot}$ is the star's equatorial rotation speed. 
Last, the magnetisation of the stellar winds is treated as a 
Parker spiral. It is a split monopole, 
\begin{equation}
	B_{\rm r}(r) = B_{\star} \Big( \frac{R_{\star}}{r} \Big)^{2},
    \label{eq:Br}
\end{equation}
where $R_{\star}$ is the stellar radius, $B_{\star}$ the stellar surface 
magnetic field, plus a toroidal component, 
\begin{equation}
	B_{\phi}(r) = B_{\rm r}(r)
	\Big( \frac{ v_{\phi}(\theta) }{ v_{\rm w} } \Big) 
	\Big( \frac{ r }{ R_{\star} }-1 \Big),
\label{eq:Bphi}
\end{equation}
which relies on both the stellar surface magnetic field strength and 
on the stellar rotation velocity $v_{\rm rot}$. This so-called Parker 
stellar wind is a parametrisation for magnetised low-mass stars which 
has been developed and widely-used in studies devoted to the  
heliosphere~\citep{parker_paj_128_1958,pogolerov_aa_321_1997,
pogolerov_aa_354_2000,pogolerov_apj_614_2004}. 
It has been adapted to the intermediate stellar mass regime in 
works investigating the shaping of planetary nebula~\citep{chevalier_apj_421_1994,
rozyczka_apj_469_1996,garciasegura_apj_860_2018,garciasegura_apj_893_2020} 
and stellar wind bow shocks~\citep{herbst_apj_897_2020,scherer_mnras_493_2020,
2021arXiv210705513M}, respectively.

\subsection{Time-dependent stellar boundary conditions}
\label{sect:bc}

At time $t_{\rm wr}-\Delta t$ the blue supergiant to Wolf-Rayet phase 
begins \textcolor{black}{to take} place, controlled by several quantities, namely its 
duration $\Delta t$ and the manner the surface stellar properties \textcolor{black}{evolve}, 
determined by the quintuplet ($\dot{M}$,$v_{\rm w}$,$v_{\rm rot}$, 
$B_{\star}$,$R_{\star}$). 
The stellar wind velocity is modulated following~\citep{brighenti_mnras_285_1997},
\begin{align}
   v_{\rm w}(r_{\rm in},t) = \Bigg\{
             \begin{array}{lll}
    v_{\rm w}^{\rm bsg} / \sqrt{ f(\theta) }             & \mathrm{if}    &  t \le t_{\rm wr}-\Delta t , \\
    v_{\rm w}^{\rm bsg}  / \sqrt{ f(\theta) }+\Delta vF  & \mathrm{if}    &  t_{\rm wr}-\Delta t < t and t \le t_{\rm wr}, \\
    v_{\rm w}^{\rm wr}                                   & \mathrm{if}    &  t_{\rm wr} < t, \\             
             \end{array}
\label{eq_vel}             
\end{align}
with $\Delta v = |v_{\star}^{\rm wr} - v_{\star}^{\rm bsg}|$. In the above and 
following relations, the superscript "bsg" and "wr" stand for the blue supergiant 
and Wolf-Rayet winds, respectively. The manner the wind velocity changes over $\Delta t$ 
is controlled by, 
\begin{equation}
	F = F(t,t_{\rm wr},\Delta t) = \Big( \frac{ t-(t_{\rm wr}-\Delta t) }{ \Delta t } \Big)^{\beta/2},
    \label{eq:F}
\end{equation}
that is a function invoked when $t_{\rm wr}-\Delta t < t$ and $t \le t_{\rm wr}$, 
defined such that $F(t=t_{\rm wr}-\Delta t)=0$ and $F(t=t_{\rm wr})=1$. 
For $\beta=2$ the transition is linear, while it is a power-law for $\beta\neq2$. 
The changes in wind density are as follows,
\begin{align}
   \dot{M}(r_{\rm in},t) = \Bigg\{
             \begin{array}{lll}
             \dot{M}^{\rm bsg} & \mathrm{if} &    t \le t_{\rm wr}, \\
             \dot{M}^{\rm wr}  & \mathrm{if} &    t_{\rm wr}< t,    \\
             \end{array}
\label{eq_mdot}               
\end{align} 
with a sharp transition at $t_{\rm wr}$~\citep{brighenti_mnras_285_1997}. 
These relations implies that only the Wolf-Rayet \textcolor{black}{wind} is launched 
isotropically while the previous winds are affected by the isotropy function $f(\theta)$.

Our MHD setup requires additional, similar prescriptions for the other quantities 
necessary to fully fix both the Parker spiral and the passive tracers 
tracking the mixing of material into the gas nebula. 
We introduce, in the same fashion as done for the hydrodynamical case 
of~\citet{brighenti_mnras_285_1997}, prescriptions for the toroidal and poloidal 
components of the stellar surface magnetic field. They read, 
\begin{align}
   B_{\star}(r_{\rm in},t) = \Bigg\{
             \begin{array}{lll}
    B_{\star}^{\rm bsg}   & \mathrm{if}     &    t \le t_{\rm wr}-\Delta t , \\
    B_{\star}^{\rm bsg} + \Delta B_{\star}F & \mathrm{if}    & t_{\rm wr}-\Delta t < t and t \le t_{\rm wr}, \\
    B_{\star}^{\rm wr}    & \mathrm{if}     &    t_{\rm wr} < t, \\             
             \end{array}
\end{align}
with $\Delta B_{\star} =| B_{\star}^{\rm wr} - B_{\star}^{\rm bsg}| $. 
Equivalently, the stellar radius evolves as, 
\begin{align}
   R_{\star}(r_{\rm in},t) = \Bigg\{
             \begin{array}{lll}
             R_{\star}^{\rm bsg}                           & \mathrm{if} &    t \le t_{\rm wr}-\Delta t , \\
             \Delta R_{\star}(1-F)  +  R_{\star}^{\rm wr}  & \mathrm{if} & t_{\rm wr}-\Delta t < t and t \le t_{\rm wr}, \\
             R_{\star}^{\rm wr}                            & \mathrm{if} &    t_{\rm wr} < t, \\             
             \end{array}
\label{Eq_radius}             
\end{align}
with $\Delta R_{\star}=|R_{\star}^{\rm bsg}-R_{\star}^{\rm wr}|$. Finally, the time-dependent 
evolution of the equatorial angular velocity reads, 
\begin{align}
   v_{\rm rot}(r_{\rm in},t) = \Bigg\{
             \begin{array}{lll}
             v_{\rm rot}^{\rm bsg}                           & \mathrm{if} &    t \le t_{\rm wr}-\Delta t , \\
             \Delta v_{\rm rot}(1-F)  +  v_{\rm rot}^{\rm wr}  & \mathrm{if} & t_{\rm wr}-\Delta t < t and t \le t_{\rm wr}, \\
             v_{\rm rot}^{\rm wr}                            & \mathrm{if} &    t_{\rm wr} < t, \\             
             \end{array}
\end{align}
with $\Delta v_{\rm rot}=|v_{\rm rot}^{\rm bsg}-v_{\rm rot}^{\rm wr}|$. 
Last, the evolution of \textcolor{black}{the} passive scalar tracer at the stellar wind boundary 
obeys the relation, 
\begin{align}
   Q(r_{\rm in},t) = \Bigg\{
             \begin{array}{lll}
             0   & \mathrm{if} &    t \le t_{\rm wr}-\Delta t , \\
             F   & \mathrm{if} &   t_{\rm wr}-\Delta t < t and t \le t_{\rm wr}, \\
             1   & \mathrm{if} &    t_{\rm wr} < t, \\             
             \end{array}
\label{Eq_tracers}              
\end{align}
which permits to separate the cold, dusty, blue supergiant gas ($Q\le1/2$) 
from the Wolf-Rayet material ($Q>1/2$).

\begin{figure*}
        \centering
        \includegraphics[width=0.75\textwidth]{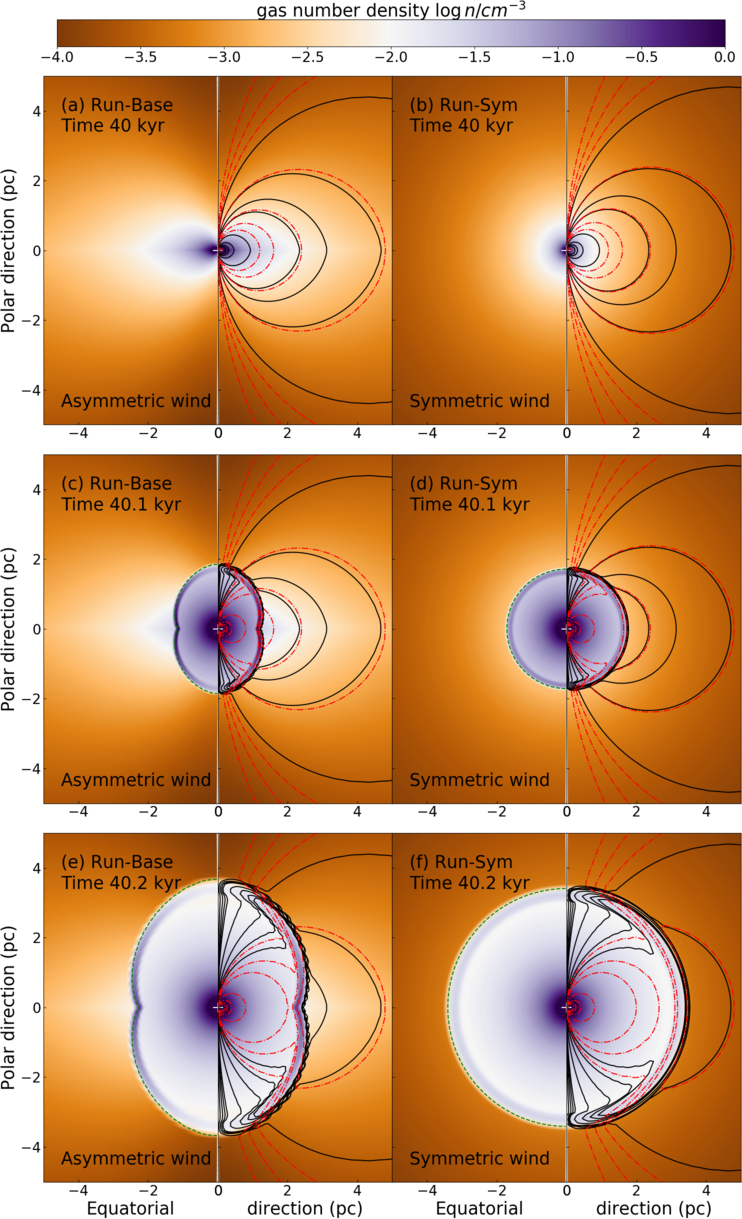} 
        \caption{
        Number density field in our simulation models Run-Base (a) 
        and Run-Sym (b).  
        The dashed green contour is the tangential discontinuity of the Wolf-Rayet nebula, 
        i.e. the location of the nebula made of equal proportion of 
        blue supergiant and Wolf-Rayet stellar winds, where the passive scalar $Q=0.5$. 
        The solid black and dashed-dotted red contours on the right-hand part 
        of each panels represent isovalues of the toroidal component of the 
        velocity $v_{\phi}$ and magnetic field $B_{\phi}$, respectively. 
        The white cross marks the position of the star. 
        }
        \label{fig:simulation1}  
\end{figure*}

\begin{figure*}
        \centering
        \includegraphics[width=1.0\textwidth]{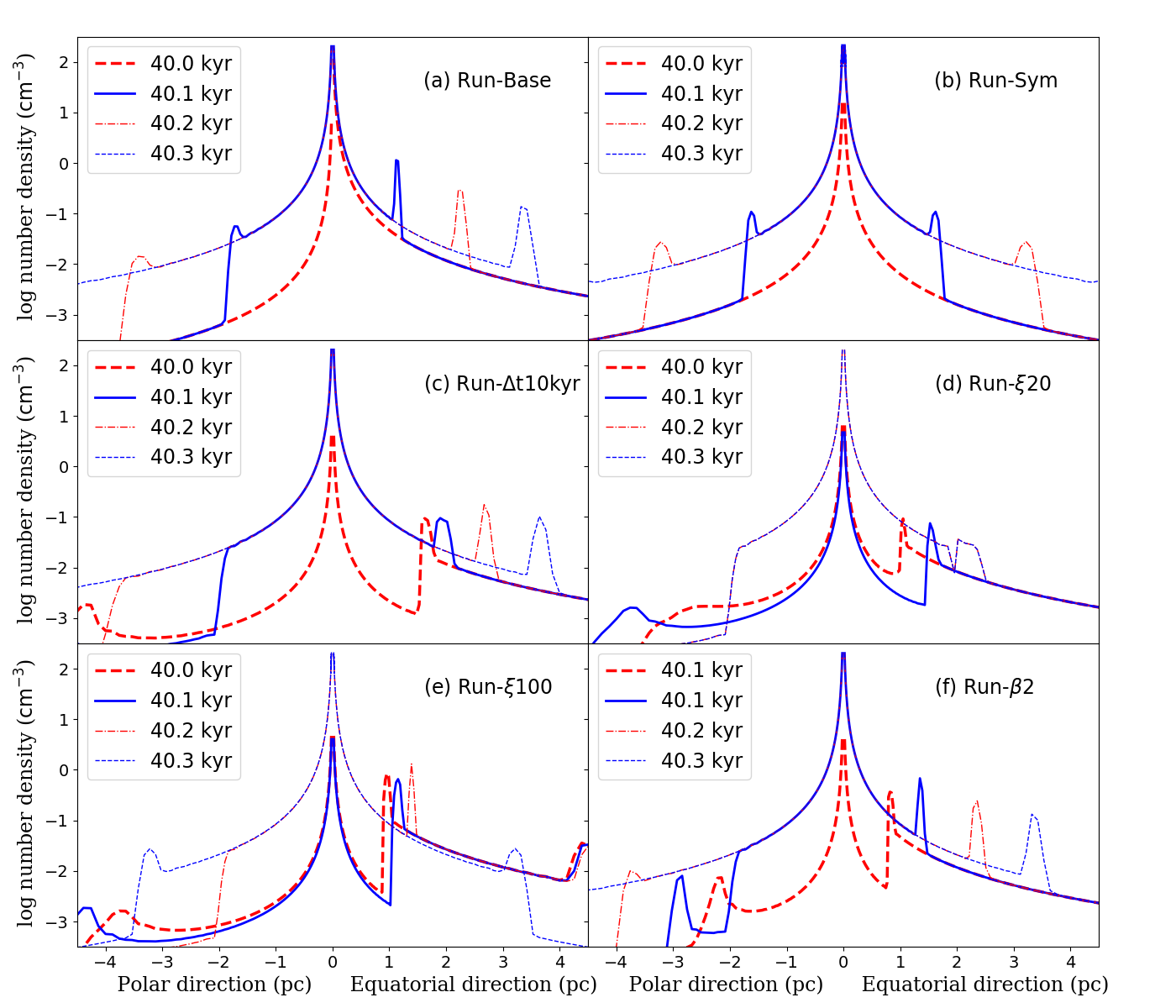} 
        \caption{
        Cross-sections in the density field and toroidal magnetic field 
        of the evolving Wolf-Rayet nebula. The panels correspond to 
        different simulations, and on each panel the cross-section are 
        plotted for both directions along the polar (left) and 
        equatorial (right) directions of the domain. 
        }
        \label{fig:section1}  
\end{figure*}

\begin{figure*}
        \centering
        \includegraphics[width=0.75\textwidth]{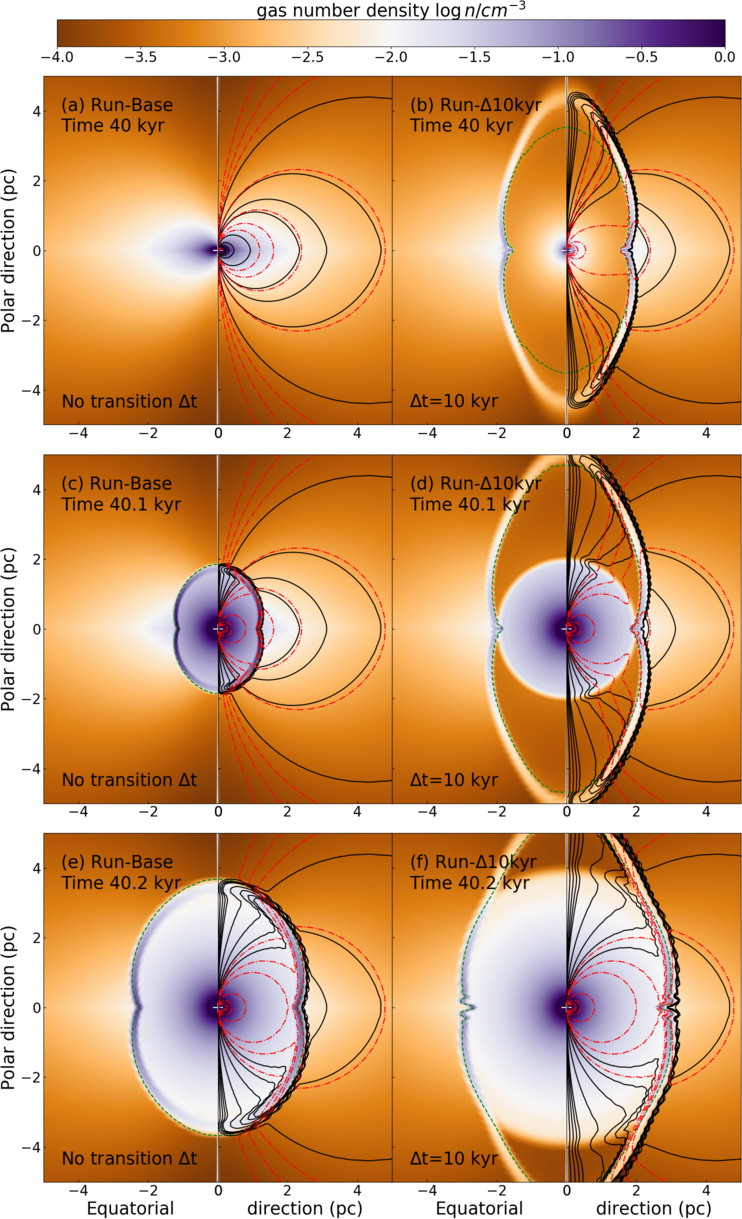} 
        \caption{
        Same as Fig.~\ref{fig:simulation1} with our models Run-Base 
        and Run-$\Delta$t10kyr, exploring the effect of a phase 
        transition timescale $\Delta t \neq 0$.
        }
        \label{fig:simulation2}  
\end{figure*}

\subsection{Simulation models}
\label{sect:models}

The simulations are initialised with stellar wind properties corresponding 
to a blue supergiant star. We let the system evolved \textcolor{black}{to} a few 
$10^{4}\, \rm yr$ up to \textcolor{black}{a} time $t_{\rm wr}-\Delta t$. This time interval corresponds 
to the typical timescale of a luminous blue variable phase~\citep{smith_rspta_2017} 
and it \textcolor{black}{permits} to have potential magneto-hydrodynamical boundary effects provoked by 
the onset of stellar rotation at $t=0$ to be transported out of the 
computational domain under the effects of \textcolor{black}{the} constant wind inflow. 
The blue supergiant stellar properties are as follows. We use 
$\dot{M}^{\rm bsg}=10^{-6}\, \rm M_{\odot}\, \rm yr^{-1}$, which is of 
the order of magnitude of that of Sher~25~\citep{smartt_aa_391_2002}, 
$v_{\rm w}^{\rm bsg}=500\, \rm km\, \rm s^{-1}$, that is consistent with 
the values for HD 168625~\citep[$350 \pm 100\, \rm km\, \rm s^{-1}$,][]{mahy_aa_594_2016} 
and for Sk$-$69\degree279~\citep[$800 \pm 100\, \rm km\, \rm s^{-1}$,][]{gvaramadze_mnras_474_2018}.
The stellar radius is $R_{\star}^{\rm bsg}=20\, \rm R_{\odot}$, which is that of 
$Sk-69\degree279$~\citep[$800 \pm 100\, \rm km\, \rm s^{-1}$,][]{gvaramadze_mnras_474_2018}, 
and $v_{\rm rot}^{\rm bsg}=60\, \rm km\, \rm s^{-1}$~\citep{mahy_aa_594_2016}. 
The surface magnetic field of the blue supergiant ancestor star is taken to \textcolor{black}{be}
$B_{\star}^{\rm bsg}=1\, \rm G$, which is of \textcolor{black}{the order of the magnetic fields 
observed in other} cool stars~\citep{kervella_aa_609_2018,vlemmings_aa_394_2002,vlemmings_aa_434_2005}.

The toroidal component of the stellar magnetic field is scaled with 
\textcolor{black}{the value for the solar} wind measured at $1\, \rm 
au$~\citep{herbst_apj_897_2020,scherer_mnras_493_2020}. 
The Wolf-Rayet stellar properties are as follows. We use 
$\dot{M}^{\rm wr}=10^{-4.3}\, \rm M_{\odot}\, \rm yr^{-1}$, 
$v_{\rm w}^{\rm wr}=1900\, \rm km\, \rm s^{-1}$, 
$R_{\star}^{\rm wr}=2.3\, \rm R_{\odot}$ and 
$v_{\rm rot}^{\rm wr}=10\, \rm km\, \rm s^{-1}$, respectively. 
These quantities \textcolor{black}{are the values} of the Wolf-Rayet 
star WR1 (WN4). Our stellar parameters are summarised in 
Table~\ref{tab:stars}. 
The stellar surface magnetic field is rather unconstrained, and we 
decided to take $B_{\star}^{\rm wr}=100\, \rm G$. 
Similarly, we assume that the toroidal component of the magnetic field 
scales with the magnetic field in the solar 
wind~\citep{herbst_apj_897_2020,scherer_mnras_493_2020}.

In this work, we use our toy model to explore the effects of the 
progenitor stellar wind and phase transition properties 
\textcolor{black}{on the shaping} of the Wolf-Rayet nebulae. 
The simulation models are summarized in Table~\ref{tab:models}.

\begin{figure*}
        \centering
        \includegraphics[width=0.75\textwidth]{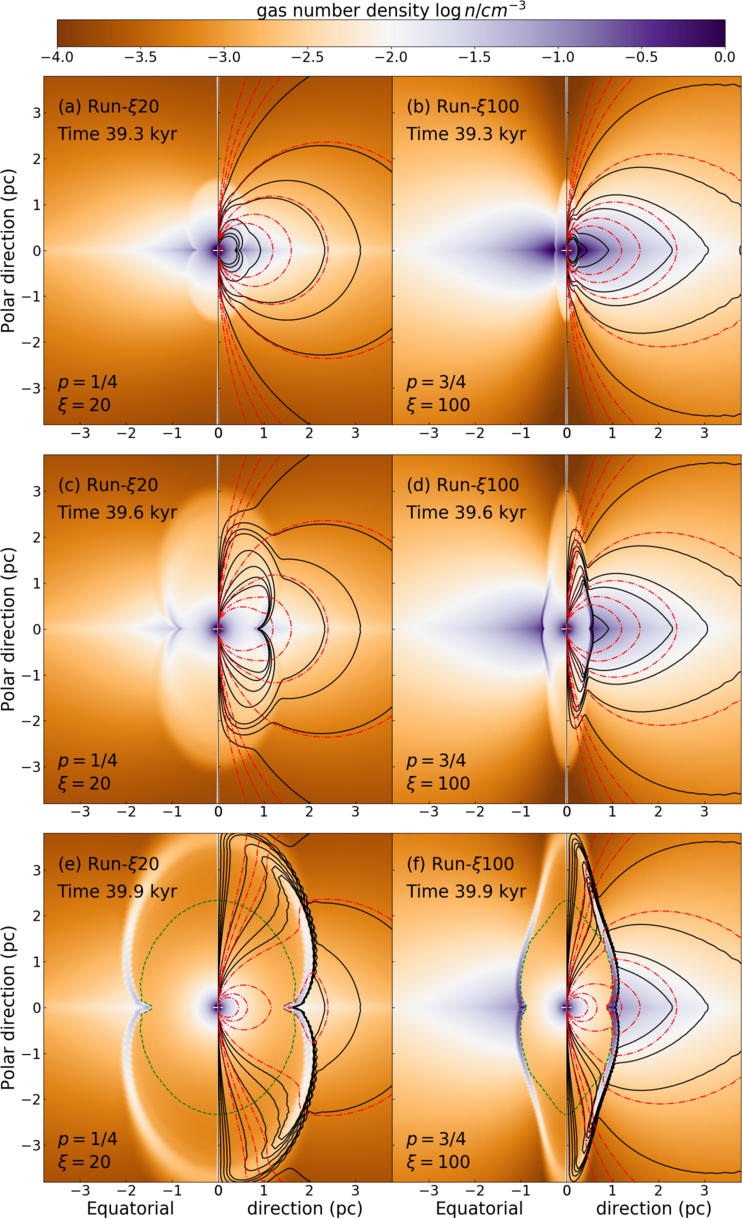} 
        \caption{
        Same as Fig.~\ref{fig:simulation1} with our models Run-$\xi$20 
        and Run-$\xi$100, exploring the effect of the flattening degree $p$  
        and density ratio $\xi$ of the progenitor wind.
        }
        \label{fig:simulation3}  
\end{figure*}

\subsection{Radiative transfer calculations}
\label{sect:rt}

\textcolor{black}{As most} circumstellar nebulae around evolved massive 
stars \textcolor{black}{have} been observed in 
the (near)-infrared waveband, we consequently desire, in order 
to discuss our models in the context of real data, to know how our simulated  
objects would look like if observed at that particular waveband. 
We perform radiative transfer calculations of our MHD simulation outputs against 
dust opacity using the {\sc radmc-3d} code~\citep{dullemond_2012}. Dust density fields 
are constructed by assuming a standard gas-to-dust ratio of $1/200$ and it is imported 
into {\sc radmc-3d}. The dust temperature is simulated by Monte-Carlo calculation 
using the algorithm of~\citet{bjorkman_paj_554_2001} and photon packages are ray-traced 
from the stellar atmosphere to the stellar wind nebulae. The $24\, \mu \rm m$ near-infrared 
emissivity is estimated using the Silicates dust opacity of~\citet{laor_apj_402_1993} 
and it is projected onto the plane of the sky according to a particular selected 
viewing angle. 
The irradiating central massive star is assumed to be a spherical black body 
radiator of effective temperature $T_{\rm eff}$ that is taken as, 
\begin{align}
   T_{\rm eff}(t) = \Bigg\{
             \begin{array}{lll}
             T_{\rm eff}^{\rm bsg}                           & \mathrm{if} &    t \le t_{\rm wr}-\Delta t , \\
             \Delta T_{\rm eff}(1-F)  +  T_{\rm eff}^{\rm wr}  & \mathrm{if} & t_{\rm wr}-\Delta t < t and t \le t_{\rm wr}, \\
             T_{\rm eff}^{\rm wr}                            & \mathrm{if} &    t_{\rm wr} < t, \\             
             \end{array}
\end{align}
where $\Delta T_{\rm eff}=|T_{\rm eff}^{\rm wr}-T_{\rm eff}^{\rm bsg}$|. 
The stellar radius $R_{\star}(t)$ is determined from Eq.~\ref{Eq_radius}. 
The hot, Wolf-Rayet material is distinguished from the cold blue supergiant 
gas using the passive scalar tracer of Eq.~\ref{Eq_tracers}. 
We produce emission maps for a distance to the stellar source of $1\, \rm kpc$, 
that is typical for massive star-forming regions.


\section{Results}
\label{sect:results}

This section presents the results \textcolor{black}{from} the modelling of Wolf-Rayet gas nebulae of blue 
supergiant ancestors. We compare the effects of the asymmetries and evolutionary phases 
of various pre-Wolf-Rayet stellar winds \textcolor{black}{on} the geometry of the gas nebulae.

\subsection{Wolf-Rayet nebulae of asymmetric versus spherical blue supergiant wind}
\label{sect:wind_asym}

In Fig.~\ref{fig:simulation1} we plot the number density field in our baseline simulation 
Run-Base (left column of panels) and our model Run-Sym (right column of panels),   
which assume asymmetric and isotropic blue supergiant progenitor stellar winds, respectively. 
The dashed green contour is the tangential discontinuity, where blue supergiant and Wolf-Rayet 
stellar winds are in equal proportions ($Q=0.5$). Solid black and dashed-dotted red contours 
trace $v_{\phi}$ and $B_{\phi}$, respectively, and the white cross marks the position of the star. 
The model Run-Sym obviously generates a spherical nebula (Fig.~\ref{fig:simulation1}d,f) 
by fast-wind-slow-wind collision, as modelled in the work of~\citet{meyer_mnras_496_2020}. 
In model Run-Base the asymmetric blue wind is expelled as a torus of outflowing material, 
see the studies of~\citet{raga_apj_680_2008} and~\citet{fang_mnras_464_2017}, 
respectively, onto which stellar rotation and a Parker wind is superposed 
(Fig.~\ref{fig:simulation1}a). 
Both models have no phase transition, therefore, the Wolf-Rayet stellar wind 
directly interacts with the luminous blue supergiant stellar wind. 
The initial spherically-symmetric stellar wind of Run-Base is channelled by its 
surroundings and adopt\textcolor{black}{s} a bipolar morphology (Fig.~\ref{fig:simulation1}c,e). 
One clearly sees magnetic dipoles in both the old slow and new fast winds, 
whose field lines are compressed in the shell region.  
Given the size of the nebula ($\le 10\, \rm pc$), the asymmetric nebulae can 
develop inside of the region bordered by the termination shock of the main-sequence 
stellar wind bubble~\citep{meyer_mnras_496_2020}. These rings should therefore form 
even in the context of the supersonic bulk motion of the driving blue supergiant star  
(Fig.~\ref{fig:simulation1}b,d,f). 
This mechanism is therefore the MHD equivalent of the young aspherical Wolf-Rayet nebulae 
of red supergiant stars of~\citet{brighenti_mnras_285_1997}, this time in the context 
of a blue supergiant ancestor.

Fig.~\ref{fig:section1} shows cross-sections taken through the number density field 
of the Wolf-Rayet nebulae. 
The figure plots the cross-sections, taken \textcolor{black}{at} intervals of $0.1\, \rm kyr$ through 
several characteristic simulation snapshots of the blue supergiant to Wolf-Rayet phase 
transition event. 
The left-hand part of each panel corresponds to the polar direction, whereas the right-hand 
part of the \textcolor{black}{panels} corresponds to the equatorial plane of the nebulae. 
The spherical nebula in Run-Sym (Fig.~\ref{fig:section1}b) is obviously similar 
along both the polar and equatorial directions. The shell of swept-up blue wind 
grows under the effect of the expelled Wolf-Rayet momentum and the contrast in 
density along both direction 
decreases, i.e. the post-shock density of the Wolf-Rayet nebulae diminishes with 
time and so does the corresponding emission (see Section~\ref{sect:emission_maps}).
\textcolor{black}{
At time $40.1\, \rm kyr$, the cross-sections of the asymmetric model Run-Base 
exhibit clear differences between the polar and equatorial directions 
(Fig.~\ref{fig:section1}a), as a result of the wind asymmetry. 
}
When the Wolf-Rayet material begins to blow, the shell 
of swept-up supergiant gas is denser in the equatorial plane than in 
the direction perpendicular to the equator, see thick blue and thin lines of 
Fig.~\ref{fig:section1}a. This latitude-dependent density contrast is at the 
origin of the asymmetric character of some observed Wolf-Rayet nebulae.

\begin{figure*}
        \centering
        \includegraphics[width=0.75\textwidth]{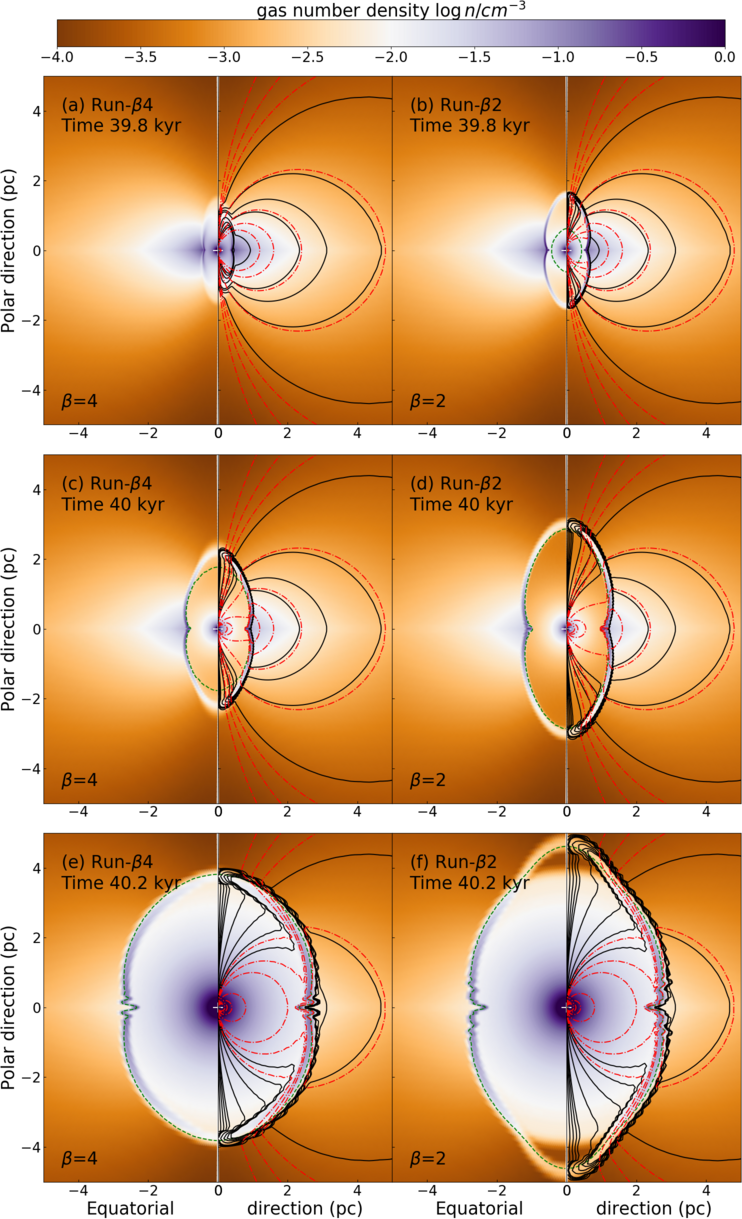} 
        \caption{
        Same as Fig.~\ref{fig:simulation1} with our models Run-$\beta$4
        and Run-$\beta$2, exploring the effects of the interpolation exponent $\beta$ 
        during the phase transition. 
        }
        \label{fig:simulation4}  
\end{figure*}

\subsection{Effects of the phase transition timescale}
\label{sect:timescale}

Fig.~\ref{fig:simulation2} compares the density fields in our simulation models 
Run-Base (left panels) and Run-$\Delta$t10kyr (right panels), which differs by the blue 
supergiant to Wolf-Rayet phase transition interval. Such transition period is non-existent  
in model Run-Base while it is $\Delta\,t=0.1\, \rm kyr$ in model Run-$\Delta$t10kyr. 
Our phase transition timescale is in accordance with both the typical duration of 
luminous blue phases $\sim 10^{4}\, \rm yr$ and with the timescale of blue-to-red 
phase transition~\citep{mackey_apjlett_751_2012,groh_aa564_2014}. 
The Wolf-Rayet nebula in Run-Base is generated by the simultaneous changes of 
the wind velocity and density, engendering a peanut-like morphology, which shocked layer 
accumulates compressed magnetic field lines, as described in 
Section~\ref{sect:wind_asym}. 
In Run-$\Delta$t10kyr the stellar wind velocity increases first, provoking a more elongated 
jet-like nebula, before that the sudden changes in wind density fills it with denser, isotropic   
Wolf-Rayet material (see Eq.~\ref{eq_vel}-\ref{eq_mdot}). In the latter case, Rayleight-Taylor 
instabilities form in the equatorial plane, as a consequence of the wind-wind collision 
at work therein, see \textcolor{black}{the} discussion in~\citet{brighenti_mnras_285_1997}. 
This comparison shows that the properties of the time windows during which the wind properties 
evolve has a little influences \textcolor{black}{on} the overall appearance of Wolf-Rayet nebulae. 
Our simplistic setup uses analytic prescriptions that mimic the evolution of 
the central massive star, hence, the detailed shape of the nebula might be 
slightly different with more realistic, time-dependent stellar wind prescriptions, 
e.g. using pre-computed evolutionary tracks~\citep{brott_aa_530_2011a}.

In model Run-$\Delta$t10kyr, the accelerating stellar wind has 
already interacted with the dense equatorial blue supergiant wind, resulting 
in a bipolar shell (Fig.~\ref{fig:simulation2}a,b) that is later filled by 
the Wolf-Rayet wind. 
The wind-wind interaction is stronger for a reduced phase transition interval $\Delta t$ 
(Fig.~\ref{fig:simulation2}a,c), while the density of the shocked material in 
Run-$\Delta$t10kyr is reduced, as a consequence of the diminished wind momentum 
at the moment just before the sudden changes in mass-loss rate at $t_\mathrm{wr}$.
Nevertheless, in both models, the Wolf-Rayet wind eventually melts with the shell of 
swept-up blue supergiant material, and, comparable equatorial densities are reached 
(Fig.~\ref{fig:simulation2}a) as it expands outwards (Fig.~\ref{fig:simulation2}c).
Finally, once the hot gas filled the bipolar cavity, both nebulae have similar 
overall dimensions except \textcolor{black}{at the poles, where  
the magnetic field lines are collimated} (Fig.~\ref{fig:simulation2}d). 
This further illustrates that the morphology of older asymmetric Wolf-Rayet nebulae
is governed by the surface characteristics and stellar wind properties, such as terminal 
velocity and mass-loss rate of the ancestor stars, as well as \textcolor{black}{the 
manner in which the} evolutionary phase transition happens, but not by the ambient 
medium properties.

\subsection{Effects of the blue supergiant stellar wind asymmetries}
\label{sect:wind_kind_asym}

Fig.~\ref{fig:simulation3} compares two simulation models, Run-$\xi$20 and Run-$\xi$100, 
both with the same phase transition timescale $\Delta t=10\, \rm kyr$, but different 
density flattening degrees towards the equator $p$ and density equator-to-pole 
ratios $\xi$, namely, $p=1/4$ and $\xi=20$ (Run-$\xi$20) and $p=3/4$ and $\xi=100$
(Run-$\xi$100), respectively. 
The different pre-Wolf-Rayet morphologies of the circumstellar medium profoundly impact the 
development of the shell generated by wind-wind interaction between blue supergiant 
to the Wolf-Rayet winds. 
The denser the material in the equator, the more collimated the Wolf-Rayet stellar 
wind (Fig.~\ref{fig:simulation3}b), as a result of the thicker disc-like blue 
supergiant distribution (Fig.~\ref{fig:simulation3}a). On the other hand, 
the simulation with a thinner equatorial plane produces a peanut-like nebula, 
recalling the mechanisms for the shaping of the homunculus of 
$\eta$-Carina~\citep{langer_ApJ_520_1999,gonzales_apj_616_2004, hirai_mnras_503_2021}. 
Note the gradual opening of the magnetic field lines as the wind speed 
accelerates (Fig.~\ref{fig:simulation2}a-e,b-f). 
Again, this is mostly shaped by the blue supergiant \textcolor{black}{wind whose velocity} 
gradually increases during the phase transition up to reaching the fast 
values of the Wolf-Rayet star (black dotted line in Fig.~\ref{fig:simulation3}c,d). 
The discrepancy between peanut-like and jet-like Wolf-Rayet nebulae persists at later times, 
when the stellar phase transition keeps going (Fig.~\ref{fig:simulation3}e,f).

As soon the Wolf-Rayet stellar wind interacts with the walls of the nebula, the 
wind-wind interface becomes naturally much denser in the model with $\xi=100$ 
than in the case with $\xi=20$ (Fig.~\ref{fig:simulation3}c,d).   
This reflects in the cross-sections taken through the density fields of the  
simulations (Fig.~\ref{fig:simulation3}d,e). Indeed, the model with $\xi=20$ 
produces a larger shell in the equatorial plane, whereas that with $\xi=100$ 
is more compact, thinner and also much denser. 
The density jump at the expanding front is larger at time $40.1\, \rm kyr$ 
in our model Run-$\xi$100, than in our simulation Run-p1/2 and it stalls 
at $\approx 1\, \rm pc$ from the star, whereas the other nebula \textcolor{black}{in Run-p1/2}
is more enlarged in the equatorial plane. 
The flattening degree of the density towards the equator and the equator-to-pole 
density ratio turns to be a major parameter in the shaping of Wolf-Rayet nebulae, 
as it controls their final aspect ratio and therefore their overall morphology. 
Within the explored parameter space, we therefore produce either peanut-like or 
jet-like shapes.

\subsection{Effects of phase transition properties}
\label{sect:transition}

Our last series of comparison tests consist in changing the interpolation 
between the blue supergiant and Wolf-Rayet stellar surface properties during 
the phase transition. It is governed by the exponent parameter $\beta$ in 
Eq.~\ref{eq:F}, \textcolor{black}{set} to $\beta=2$ in our model Run-$\beta$2 (linear interpolation) 
and to $\beta=4$ in our Run-$\beta$4 (squared polynomial interpolation). 
The transition phase timescale is set to $\Delta t=5\, \rm kyr$. 
In Fig.~\ref{fig:simulation4}a and Fig.~\ref{fig:simulation4}b, the pre-shaped 
circumstellar medium in which the Wolf-Rayet star blows its material is similar, 
and how the blue supergiant stellar wind gradually adopts a faster speed 
is the only effect responsible for differences in the nebula's morphologies. 
This keeps on going for another $0.1\, \rm kyr$, until the tangential 
discontinuity between the two colliding winds reach\textcolor{black}{es} the cavity. 
Nevertheless, when the full Wolf-Rayet wind blows through and fully fills the 
cavity, morphological differences are rather small (Fig.~\ref{fig:simulation4}e,f). 
The field lines are opened under the effect of the faster equatorial rotation 
of the Wolf-Rayet star. 
Instabilities developing in the equatorial plane, as in model Run-$\Delta$t10kyr, 
appear in both runs, despite of the different, longer phase transition. 
The apex of the polar lobes remains the only trace of the different 
stellar evolution history of the final Wolf-Rayet stars.

\subsection{Magnetic properties of the Wolf-Rayet nebulae}
\label{sect:magnetisation}

\begin{figure}
        \centering
        \includegraphics[width=0.45\textwidth]{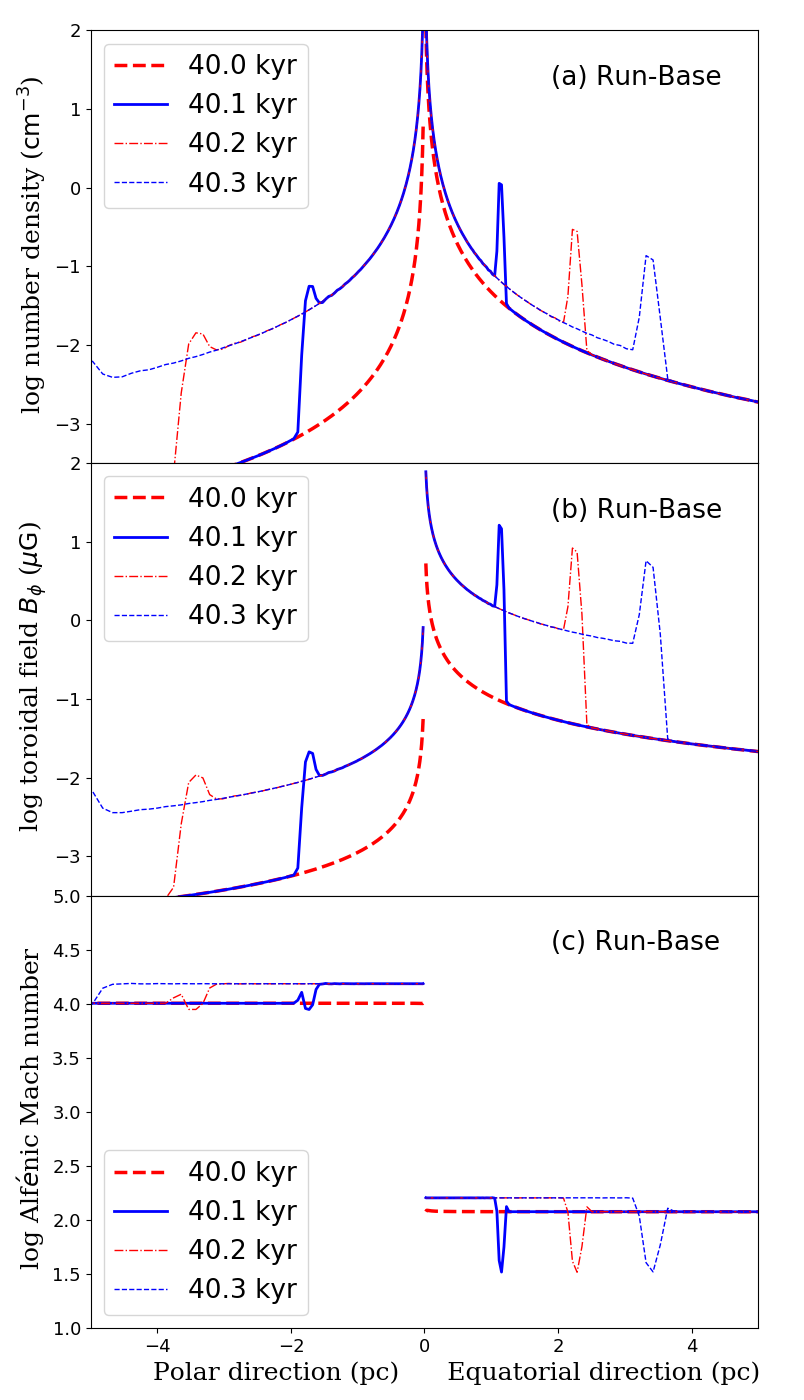} 
        \caption{
        Cross-sections in the density field (top panel), toroidal magnetic field 
        (middle panel) and Alf\'enic Mach number (bottom panel ) of the evolving 
        Wolf-Rayet nebula in our simulation Run-Base. 
        On each panel the cross-section are plotted for both the polar (left)  
        and the equatorial direction (right) of the computational domain. 
        }
        \label{fig:simulation5}  
\end{figure}

In Fig.~\ref{fig:simulation5}a we plot a time evolution series of cross-sections 
taken through the number density field (top panel) in our simulation model assuming 
an aspherical blue supergiant stellar wind (Run-Base). The selected \textcolor{black}{times} 
correspond to the moment the Wolf-Rayet wind begins blowing into the blue supergiant material. 
Fig.~\ref{fig:simulation5}b (middle panel) displays cross-sections taken at the same 
time instances and location, through the toroidal density field $B_{\phi}$ (in $\mu 
\rm G$) and plotted in the logarithmic scale. 
One sees that the magnetic field in the Wolf-Rayet nebula, governed by its toroidal 
component is maximum in the region of the expanding shell. The compression factor 
of the magnetic stellar wind field reflects that of the expanding shell density 
(Fig.~\ref{fig:simulation5}a,b), and, inversely, the toroidal field $B_{\phi}$ is 
weaker along the polar direction. 
Fig.~\ref{fig:simulation5}c (bottom panels) plots the \textcolor{black}{Alf\' enic} Mach 
number in the circumstellar nebula. The gas is super-Alf\' enic everywhere, with lower values 
$M_{\rm A}\sim 1.5$ in the region of the equatorial plane, in the compressed shell, 
and it peaks \textcolor{black}{at} $M_{\rm A}\approx 4$ along the polar direction.

\begin{figure*}
        \centering
        \includegraphics[width=1.0\textwidth]{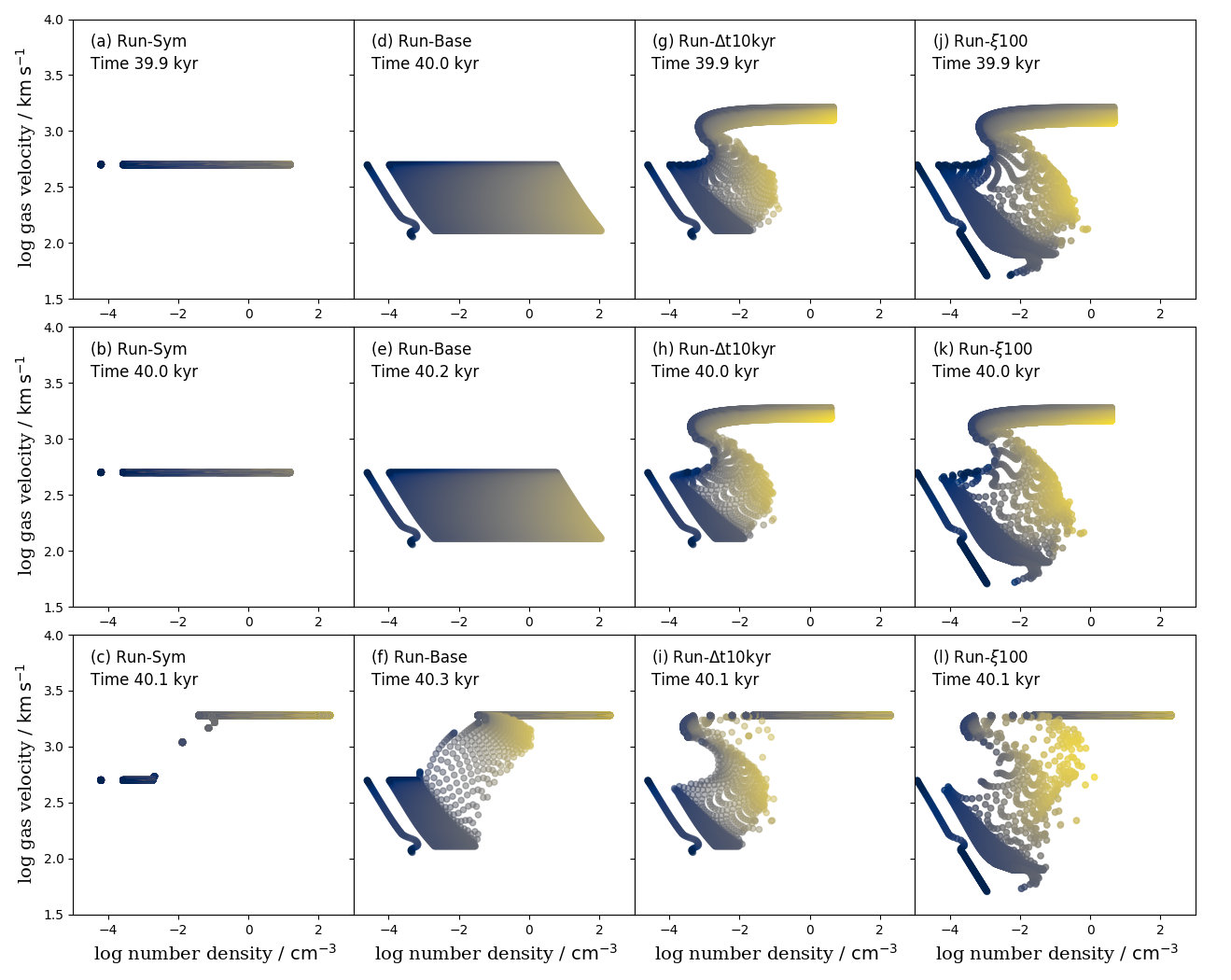} \\
        \includegraphics[width=0.8\textwidth]{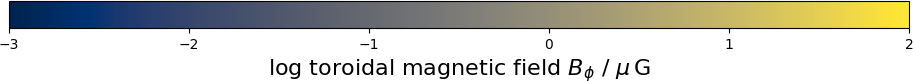} 
        \caption{
        Correlation between the number density and the velocity of the gas 
        in the inner $5\, \rm pc$ of the computational domain, for several 
        simulations of the stellar wind nebulae generated during blue supergiant 
        to Wolf-Rayet phase transitions, with colours representing the strength of 
        the toroidal of the magnetic field $B_\phi$.   
        }
        \label{fig:hist2}  
\end{figure*}

In Fig.~\ref{fig:hist2} we display scatter plots for the distribution of gas in 
the computational domain, representing the number density (in $\rm cm^{-3}$) 
of the wind as a function of its velocity (in $\rm km\, \rm s^{-1}$), 
coloured by the value of the toroidal magnetic field $B_{\phi}$ (in $\mu \rm G$). 
The different panels of the figure correspond to several time instances of the 
simulations, from the younger (top panels) to the older times (bottom panels), 
and each column represent a simulation model. 
The constant, isotropically freely-streaming stellar wind of the blue supergiant 
progenitor is represented by an horizontal line of decreasing toroidal magnetic 
field from the high to the lower densities (Fig.~\ref{fig:hist2}a,b). When 
the star evolves to the Wolf-Rayet phase, the horizontal is split into two 
horizontal lines of different velocities, a fast velocity, high density one 
(the Wolf-Rayet component), and a slow, low-density one (the blue component),  
see Fig.~\ref{fig:hist2}c.  
The model Run-Base with anisotropic stellar wind vertically spreads the horizontal line 
of Fig.~\ref{fig:hist2}a as a broadened band, \textcolor{black}{whose} height is a function 
of the \textcolor{black}{degree of flattening of the wind} and of the equator-to-pole 
density ratio, see Fig.~\ref{fig:hist2}d,e. 
The phase transition, with spherical Wolf-Rayet wind blown into an asymmetric 
blue supergiant wind, shifts the high density part of the bar-like distribution 
to a higher velocity, higher number density horizontal line (Fig.~\ref{fig:hist2}f). 
The region between the two winds exhibit\textcolor{black}{s} variations of the toroidal magnetic 
field from the density compression during the wind-wind interaction, which 
results in more scattering of the data in the velocity-density plane 
(Fig.~\ref{fig:hist2}f).

The model with both anisotropic pre-Wolf-Rayet stellar wind and long phase transition timescale, 
Run-$\Delta$t$10\rm kyr$, exhibits a distribution in the $\rho$-$v$ diagram both reporting 
the initial freely-expanding, high-density stellar wind as a 
high-density horizontal bar, plus a low-density region broadened in 
velocity, as a result of the blue supergiant asymmetries, already at work 
at that time (Fig.~\ref{fig:hist2}g). 
At later times, the transition region greatly affects the gas properties in 
the equatorial plane and this results \textcolor{black}{in a larger} toroidal magnetic 
field (Fig.~\ref{fig:hist2}h,i). At even later times, the development of 
instabilities at the wind-wind interface in the equatorial plane strongly 
disperses the high-velocity gas in the $\rho$-$v$ diagram (Fig.~\ref{fig:hist2}i). 
Similarly, the simulation with Run-$\xi$100 exhibits a more important 
scattering of the magnetised blue supergiant and expanding Wolf-Rayet stellar 
winds. The velocity broadening is strong, as a result of the large 
equator-to-pole density ratio $\xi$. The jet-like nebula has a dispersion of 
the material therein that contains highly-magnetised, swept-up, slow diluted 
blue supergiant stellar wind in the equatorial plane (Fig.~\ref{fig:hist2}k,l).  
The different velocity distributions in peanut-like ($\xi=20$) and jet-like 
($\xi=100$) Wolf-Rayet nebulae clearly appear as a consequence of the aspect 
ratio of the wind bubble, see Fig.~\ref{fig:hist2}i,l.

\begin{figure}
        \centering
        \includegraphics[width=0.47\textwidth]{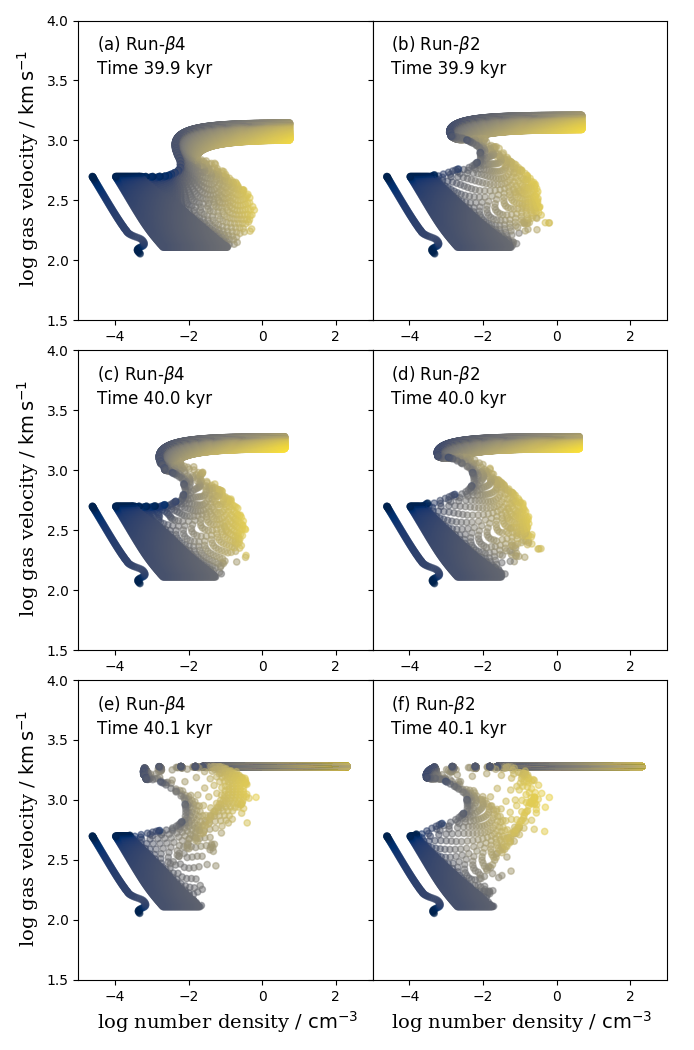} \\
        \centering
        \includegraphics[width=0.40\textwidth]{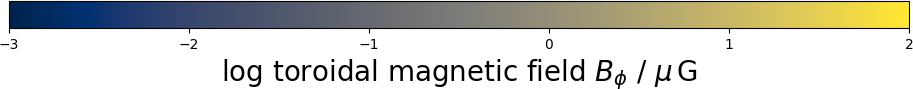} 
        \caption{
        Same as Fig.~\ref{fig:hist2} for our models Run-$\beta$4 and Run-$\beta$2.   
        }
        \label{fig:hist3}  
\end{figure}

Fig.~\ref{fig:hist3} \textcolor{black}{shows} our models Run-$\beta$4 and 
Run-$\beta$2, respectively, which differ by the manner the various quantities 
of the stellar wind are interpolated during the phase transition. 
At time $40\, \rm kyr$ both nebulae have already grown when the new-born 
Wolf-Rayet wind is blown into the accelerated blue supergiant wind, as testifies 
the S-shape in the density-velocity plane. 
The distribution of the gas therein presents both the broadened low-velocity 
component from an equatorially-asymmetric blue supergiant stellar wind and 
the horizontally-distributed component of a rapid Wolf-Rayet wind. 
Both distributions with changing $\beta$ are qualitatively similar (Fig.~\ref{fig:hist3}a,b), 
while differences at the transition between the two components appear, as the 
wind-wind interaction region at the equatorial plane is not the same. The model 
with non-linear interpolation ($\beta=4$) reveals compressed gas with a stronger 
magnetisation $B_\mathrm{\phi}$. 
The phase transition continues modifying the curved, S-shaped region of scattered 
dots between the two components of the ancestor and evolved stellar winds. Such 
curvature is more pronounced in the case with $\beta=4$ because the wind momentum 
increases faster with the polynomial interpolation than in the linear case with 
$\beta=2$, which results in an accumulation of dense, magnetised and fast material 
in the equatorial plane (Fig.~\ref{fig:hist3}c,d). 
The high velocity component turns into a thin horizontal line once the Wolf-Rayet 
wind is fully established. The instabilities developing in those peanut-like 
Wolf-Rayet nebulae are responsible for the scattering of points in the high-density, 
high-velocity parts of the diagram.

\begin{figure*}
        \centering
        \includegraphics[width=0.9\textwidth]{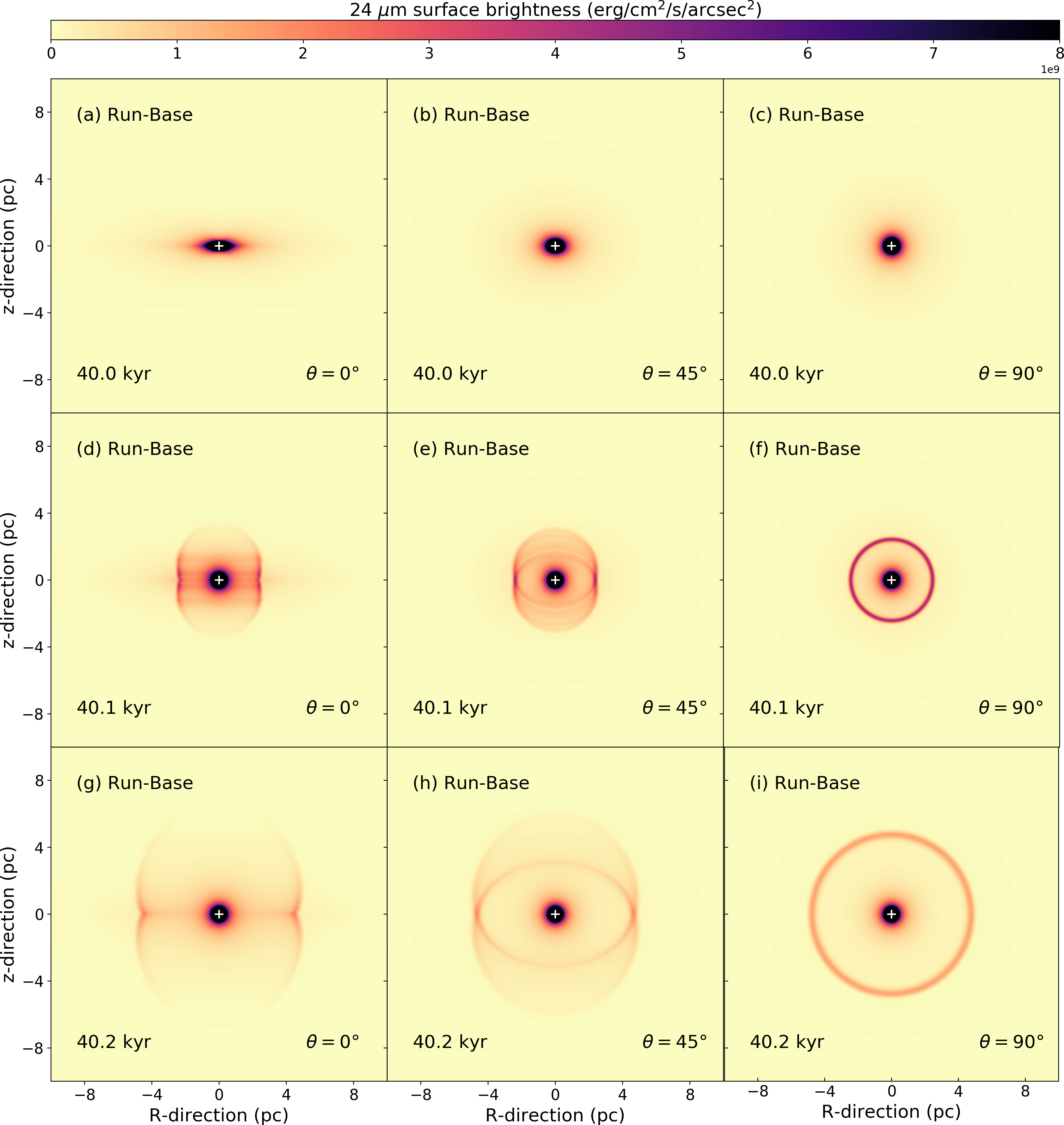} 
        \caption{
        Emission maps of calculated from the Wolf-Rayet circumstellar nebulae model Run-Base.  
        The figure plots the $24\, \mu \rm m$ surface brightness 
        (in $\rm erg\, \rm cm^{2}\, \rm s^{-1}\, \rm arcsec^{-2}$) 
        of our Wolf-Rayet nebulae. Quantities are calculated excluding the 
        non-dusty Wolf-Rayet material plotted \textcolor{black}{with} the linear scale, under 
        several viewing angles $\theta$. 
        The white cross marks the position of the star. 
        }
        \label{fig:maps1}  
\end{figure*}

\begin{figure}
        \centering
        \includegraphics[width=0.5\textwidth]{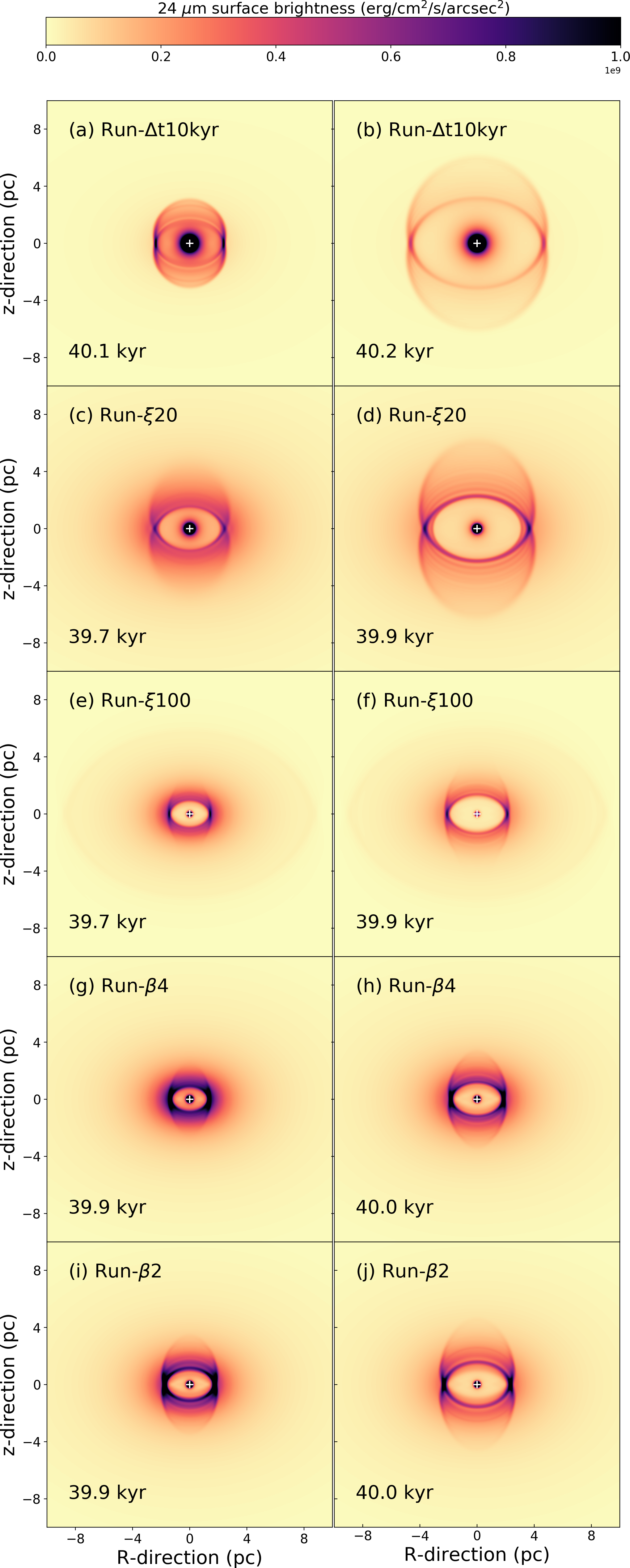} 
        \caption{
        Same as Fig.~\ref{fig:maps1} for our other models. Images assume 
        an inclination angle of $\theta=45\degree$. 
        }
        \label{fig:maps2}  
\end{figure}


\section{Discussion}
\label{sect:discussion}

This section discusses the caveats of our method and compares our 
study to results in precedent works. We also discuss the infrared emission 
properties of the young Wolf-Rayet wind nebulae, and, finally, we compare 
our results with observations.

\subsection{Model limitations}
\label{sect:caveats}

The very first caveat of our simulations is their 2.5-dimensional nature, as  
our method naturally imposes axisymmetry to the Wolf-Rayet 
nebulae~\citep{heiligman_mnras_191_1980,garciasegura_apj_517_1999}. 
While this is acceptable in our pilot study, performing full 3D MHD models 
better resolving the detailed formation of, e.g. instabilities during the 
wind-wind interaction mechanisms~\citep{brighenti_mnras_285_1997,meyer_mnras_496_2020}
should be considered in the future. 
Secondly, we neglected several microphysical processes, such as  
radiation transport in the close surroundings of the massive star. 
The high temperatures of the Wolf-Rayet star ($\sim 10^{5}\, \rm K$) must 
affect the local thermodynamical properties and the energy budget of the 
stellar wind, and, as radiation pressure is at work therein, this should 
changes the gas dynamics~\citep{2021arXiv210403968M}. 
Any chemical and/or non-ideal MHD mechanisms are also totally neglected 
in our simulations. Their consideration, much beyond the scope of our simplistic 
work, will be treated in follow-up studies. 
Last, our study which continues the work of~\citet{brighenti_mnras_285_1997}, 
utilises analytic prescriptions for the evolution of the stellar wind properties. 
It has the advantage to directly use stellar properties derived 
from e.g., observations of blue supergiant stars~\citep{smartt_aa_391_2002,mahy_aa_594_2016,
smith_rspta_2017,gvaramadze_mnras_474_2018}, however, at the cost of the introduction 
of free parameters such as the timescale $\Delta t$ of the corresponding 
phase transition. This can be circumvented by using self-consistently calculated 
theoretical stellar evolution models~\citep{brott_aa_530_2011a,2020arXiv200408203S}.

\subsection{Infrared emission maps}
\label{sect:emission_maps}

In Fig.~\ref{fig:maps1} we display $24\, \mu \rm m$ emission maps calculated on the basis of  
our Wolf-Rayet nebula model Run-Base, using the {\sc RADMC-3D} code. The surface brightness 
is plotted in the linear scale in $\rm erg\, \rm cm^{2}\, \rm s^{-1}\, \rm arcsec^{-2}$ 
for several viewing angles and the images are calculated excluding the hot dust-free 
Wolf-Rayet gas, see~\citet{meyer_mnras_496_2020}. On each panel the central white 
cross marks the position of the star. 
The first column of panels \textcolor{black}{displays the} time-dependent edge-view evolution of the 
close circumstellar medium of the massive stars. Before the evolution of the blue 
supergiant star, the asymmetric wind generates a bright oblate disc-like 
shape extending in the equatorial plane of the star (Fig.~\ref{fig:maps1}a). 
This model does not have a long phase transition interval $\Delta t$, and the hot 
Wolf-Rayet wind directly interacts with the dense equatorial colder gas. This 
interaction region emits in the mid-infrared, which produces the bipolar 
morphology (Fig.~\ref{fig:maps1}d). As the phase transition ends, the nebula 
further extends and become fainter at the poles, while conserving its overall 
bipolar bubbly shape. 
The second column of panels plots the same sequence of images considered with 
an inclination angle $\theta=45\degree$ with respect to the plane of the sky. 
The situation is qualitatively the same as for $\theta=0\degree$, except that 
the bright infrared equatorial ring appears as an ellipse (Fig.~\ref{fig:maps1}e,h). 
The last column of panels displays top-viewed nebulae, which, as a consequence of 
the 2.5D nature of the simulations, look like rings extending away from the 
evolving central star (Fig.~\ref{fig:maps1}f,i). The maximum emission region is 
the interaction layer where the Wolf-Rayet wind reaches the termination shock 
of the blue supergiant nebula (Fig.~\ref{fig:maps1}d,e,f).

Fig.~\ref{fig:maps2} is as Fig.~\ref{fig:maps1} for the other simulations models 
listed in Table.~\ref{tab:stars}. The left column of panels shows the nebula 
when the wind-wind collision begins, while the right column of panels shows the 
circumstellar nebula around the Wolf-Rayet stars at a later time. Both series of 
images are displayed with a viewing angle of $\theta=45\degree$ with respect to the 
plane of sky. 
The model Run-$\Delta$t10kyr has a phase transition \textcolor{black}{$\Delta t=10\, \rm kyr$} extending  
the time interval during which the Wolf-Rayet wind \textcolor{black}{merges} with the previous blue 
supergiant material. This results in a denser region of wind-wind interaction 
and a brighter bipolar nebula (Fig.~\ref{fig:maps2}a) materialised as an ellipsoid 
in the projected equatorial plane (Fig.~\ref{fig:maps2}b). 
Our model Run-$\xi$20 has a different density flattening degree towards the 
equator. It emphasizes the bipolar aspect of the nebula, which \textcolor{black}{is} less 
bubbly and more elongated (Fig.~\ref{fig:maps2}c,d). 
Our model Run-$\xi$100 with a larger equator-to-pole density ratio $\xi=100$  
generates a tubular nebula as during the phase transition the wind material is 
channeled into the direction perpendicular to the equator, resulting in a less 
compressed and $24\, \mu \rm m$-fainter expanding shell. 
The model Run-$\beta$4 with a non-linear interpolation of the stellar wind quantities 
during the phase transition \textcolor{black}{produces} a quicker and more violent wind-wind collision  
and a brighter interaction region (Fig.~\ref{fig:maps2}g,h). 
Our model Run-$\beta$2 with a linear interpolation induces a more important 
compression of the shell of stellar wind, and also a more pronounced bipolar 
morphology (Fig.~\ref{fig:maps2}i,j).

\subsection{Comparison with observations}
\label{sect:observations}

\subsubsection{Generalities}
\label{sect:generalities}

The geometry of Wolf-Rayet nebulae is a long-standing problem and still rather 
unexplored mystery. All except one simulation performed to date predict 
their appearance to be both spherical and affected by hydrodynamical instabilities 
originating from the evolved Wolf-Rayet stellar wind, blown into the interior of its 
own main-sequence stellar wind bubble and interacting with the shell generated during its 
previous, e.g. red supergiant, evolutionary stage~\citep{freyer_apj_594_2003,freyer_apj_638_2006, 
toala_apj_737_2011,meyer_mnras_493_2020}. All these models assume a static 
massive star undergoing a succession of isotropically-launched stellar winds. 
As discussed above, a cool, dense stellar wind launched along the equator of 
the star can make the Wolf-Rayet nebula strongly anisotropic, shaping the outflow 
as a bipolar structure. The physical origin of the equatorial 
stellar wind is so-far included as sub-grid models in the few simulations run so far, 
see~\citet{brighenti_mnras_285_1997}. Our study is therefore the first work \textcolor{black}{continuing} 
that on asymmetric Wolf-Rayet nebulae of~\citet{brighenti_mnras_285_1997}.

The previous paper of our series explored the effects of the stellar bulk 
motion \textcolor{black}{on} the formation of Wolf-Rayet stars. Indeed, the fastest runaway 
massive stars of the Milky Way \textcolor{black}{are strong-winded and bow-shock-free} Wolf-Rayet stars 
carrying circular rings, \textcolor{black}{which} challenged our understanding of the formation of 
Wolf-Rayet nebulae~\citep{meyer_mnras_496_2020}. 
A typical example is the nebula M1-67 around the star WR124~\citep{sluys_aa_398_2003}. 
The proposed simple solution for their formation scenario teaches us that the 
stellar motion of the driving star does not \textcolor{black}{influence} the growth 
of Wolf-Rayet nebulae, since the region of free-streaming between stellar surface 
and main-sequence wind bubble's termination shock is sufficiently extended. This 
is largely the case for high-latitude objects, as the diluted medium therein favors 
the production of huge and extended bow shocks, making room for wind-wind shell 
collision to happen without being \textcolor{black}{affected} by the ISM material. 
Importantly, the absence of observed ISM material in Wolf-Rayet 
nebulae~\citep{fernandezmartin_aa_541_2012,esteban_mnras_460_2016} indicates that 
most Wolf-Rayet shells, including bipolar Wolf-Rayet nebulae, are probably 
produced this way, and, this further implies that the large-scale wind-blown bubble 
can be ignored when numerically modelling them.

However, the very complex shape of many observed Wolf-Rayet nebulae challenges this 
picture~\citep{toala_aa_578_2015}. When exactly bipolarity happens and which factors 
are in control of this ? Which progenitors are responsible for which kind of circumstellar 
nebulae ? The well-studied bipolar Wolf-Rayet nebulae NGC 6888, of probable red supergiant 
ancestor, and RCW58, seemingly \textcolor{black}{coming from a} blue supergiant, are the ideal cases to test 
numerical models like ours against observations, see the discussion 
in~\citet{garciasegura_1996_aa_305f} and~\citet{garciasegura_1996_aa_316ff}. 
Bipolarity of single blue supergiant origin is the situation that we simulate in 
the present study, and it is in accordance with the interpretation 
of~\citet{garciasegura_1996_aa_316ff} in the context of RCW58. However, what 
can our result teach us regarding the other, red supergiant-based scenario ?

\subsubsection{What makes NGC 6888 asymmetric ?}
\label{sect:ngc6888}

The gaseous nebula NGC 6888, also called the veil nebula, the crescent nebula, 
Caldwell 27, Sharpless 105, is an asymmetric wind bubble found around the 
evolved massive Wolf-Rayet WN66(h)-type star WR 136. Its particular shape is 
made of two bipolar lobes and of an equatorial ring of denser material. 
\textcolor{black}{It has long been studied} in great detail~\citep{parker_apj_224_1978,treffers_apj_254_1982,garciasegura_1996_aa_316ff, 
fernandezmartin_aa_541_2012,mesadelgado_apj_785_2014,toala_aj_147_2014, 
esteban_mnras_460_2016,rubio_mnras_499_2020}. 
Early observations of the surroundings 
of NGC 6888 revealed a variation of the level ionization, traced by the 
optical line ratios [NII]/H$\alpha$ in the many clumps \textcolor{black}{along} the surface of the shell, 
witnessing an efficient mixing of materials preceding the formation of the Wolf-Rayet 
shell~\citep{parker_apj_224_1978,fernandezmartin_aa_541_2012}. The presence of dust and 
enriched elements around Wolf-Rayet nebulae indicate that (i) they result from wind-wind 
interaction of evolved material, (ii) one of the evolved \textcolor{black}{winds} is that of a cold, 
dust-producing, supergiant phase of stellar evolution and (iii) the kind of detected 
dust tells us that \textcolor{black}{the wind from a main-sequence star
does not contribute} to the shaping of the nebula. 
NGC 6888 has been the site of the first observation of the aftermath of the CNO 
cycle of nuclear reactions into the stellar core~\citep{mesadelgado_apj_785_2014},
and its progenitor star has been constrained to be a single, \textcolor{black}{red-supergiant-evolving 
star of} $25-40\, \rm M_{\odot}$~\citep{stock_mnras_441_2014,mesadelgado_apj_785_2014}. 
Hence, these \textcolor{black}{Wolf-Rayet stars} must probably have undergone a red supergiant 
evolutionary phase, and, such scenario has been modelled using 1D and 2D 
spherically-symmetric hydrodynamical \textcolor{black}{simulations} in~\citet{garciasegura_1996_aa_316ff}. 
However, if it is clear that WR136 has been a cold star prior to the Wolf-Rayet 
phase~\citep{stock_mnras_441_2014}, the origins of the progenitor's asymmetries are 
still not understood.

Our results show that a non-eruptive single blue supergiant star can not spin 
sufficiently fast to produce an asymmetric pre-Wolf-Rayet circumstellar 
medium (our Fig.~\ref{fig:simulation1}). Since 
red supergiants and other late-type stars do not rotate 
fast~\citep{kervella_aa_609_2018,vlemmings_aa_394_2002,  
vlemmings_aa_434_2005}, their shells can not be greatly affected by rotational and/or 
magnetic effects either. 
Therefore, the question is \textcolor{black}{how NGC 6888 can produce} a bipolar Wolf-Rayet nebula with an 
equatorial ring, obviously witnessing the existence of an equatorial structure produced 
during a \textcolor{black}{post-main-sequence evolutionary phase} prior to the Wolf-Rayet phase ?  
The most accepted picture, supported by CNO yields \textcolor{black}{measured} in the nebula, made of a 
main-sequence, red supergiant and Wolf-Rayet single star evolutionary channel, 
is therefore challenged by our results, as explanations for the polar asymmetries of the 
pre-Wolf-Rayet wind are still missing.

We speculate that a mechanism, similar to that revealed 
for asymptotic giant stars, see particularly~\citet{decin_sci_369_2020} 
and~\citet{2020arXiv201113472}, 
in which multiplicity \textcolor{black}{generates} a latitude dependence of the stellar wind, should 
be at work in the context of red supergiant stars as well, and in particular 
to the central star(s) of NGC 6888 before it adopted a Wolf-Rayet spectral type. 
Other possibility could be alternative evolution scenario involving an (eruptive) luminous blue 
variable event, e.g. within a blue loop~\citep{chita_aa_488_2008,mackey_apjlett_751_2012}, 
however less in accordance with the dust size measures of~\citet{rubio_mnras_499_2020} 
shown to be typical \textcolor{black}{for} red supergiants. 
Their \textcolor{black}{constraint} of the initial mass of WR136 to $\approx 50$-$60\, \rm M_{\odot}$ 
might \textcolor{black}{reconcile} both arguments, dust and asymmetries being generated at different 
phases of the complex evolution of a (potentially multiple) stellar system. 
Our allegation that the driving stellar object of NGC 6888 is not, despite the findings 
of~\citet{stock_mnras_441_2014,rubio_mnras_499_2020}, a single star, is supported by 
the detection of periodical modulation in line emission from the Wolf-Rayet star 
WR 134, interpretable as an inhomogeneous outflow~\citep{morel_apss_260_1998} 
or as a trace of a so far undetected close companion~\citep{meyer_mnras_473_2018}. 
\textcolor{black}{
The close companions to 
evolved massive stars~\citet{koenigsberger_639_aa_2020} and the circumstellar 
nebulae of extragalactic Wolf-Rayet stars such as GR290 in the Galaxy 
M33~\citep{maryeva_aa_617_2018,maryeva_aa_635_2020} also support our arguments. 
}
All this \textcolor{black}{motivates} further investigations of that problem.

\section{Conclusion}
\label{sect:conclusion}

In this paper we study the possibility of forming asymmetric Wolf-Rayet nebulae 
as a direct result of equatorial anisotropies in the blue supergiant stellar wind 
preceding the Wolf-Rayet phase. 
We present a two-winds toy model for the investigation of circumstellar medium 
generated from the evolutionary phase transition of different latitude-dependent 
distributions. 
\textcolor{black}{Our MHD simulations are performed with a 
spherically-symmetric grid} in the 2.5D fashion, using the 
{\sc pluto} code~\citep{mignone_apj_170_2007,migmone_apjs_198_2012,vaidya_apj_865_2018}. 
We concentrate on the $\sim 10\, \rm pc$ surrounding the star, and treat both stellar 
magnetic field and stellar rotation, using a combination of the recipe for asymmetric winds 
of~\citet{raga_apj_680_2008} and magnetised astropsheres~\citep{parker_paj_128_1958}. 
This study updates the pioneering work of~\citet{brighenti_mnras_285_1997} on the early 
aspherical evolution of young Wolf-Rayet nebulae from red supergiant stars. Particularly, 
our toy model permits to change the rotation, mass-loss, magnetic properties of the evolving 
stars, while simultaneously exploring the phase transition interval between the two evolutionary 
stages. It allows us to explore at reduced computational costs, the wind-wind interaction of typical 
magnetised stellar outflows responsible for the shaping of circumstellar \textcolor{black}{nebulae. 
We also explore,} by means of radiative transfer calculations, the emission properties of those gas 
nebulae.

The density distribution in the pre-Wolf-Rayet surroundings is directly responsible 
for the shaping of our Wolf-Rayet nebulae and the stellar magnetic field seems to be dynamically 
unimportant in the process of \textcolor{black}{producing} the bipolarities, in the sense 
that \textcolor{black}{only evolving MHD 
flows in rotation} produce spherical nebulae, at least in the corner of the parameter space of 
blue supergiant and subsequent Wolf-Rayet stars that we explore. 
An equatorially compressed circumstellar distribution \textcolor{black}{gives} the Wolf-Rayet stellar wind a 
peanut-like morphology, which is further elongated along the polar direction as a jet-like 
shape if the polar-to-equatorial density distribution ratio increases, or in case a 
\textcolor{black}{substantially long} phase transition period ($\sim \rm kyr$) between the two winds. 
The topology of the magnetic field lines \textcolor{black}{is greatly} modified when the Wolf-Rayet wind 
\textcolor{black}{starts} blowing into the previous cold material. The rotating blue supergiant 
dipole $B_{\phi}$ sees its magnetic field lines opened and channelled to the poles of the structure. 
Interestingly, all our nebulae are shaped by wind-wind interaction taking place in the 
free-streaming stellar wind of the ancestor star, i.e. inside of the termination shock of 
its main-sequence stellar wind bubble~\citep{weaver_apj_218_1977}. This implies that the 
nebulae will appear the same regardless of their driving star's bulk 
motion~\citep{meyer_mnras_496_2020} and that the nebulae are free of ISM 
material. 
We propose to characterise the evolution of MHD asymmetric Wolf-Rayet nebulae in 
the $\rho$-$v$-$B_{\phi}$ diagram, which we show to concisely report the main 
feature of our MHD Wolf-Rayet nebulae as a characteristic S-shape.

Radiative transfer calculations against dust opacity of the Wolf-Rayet nebulae 
provided us with mid-infrared $24\, \mu  \rm m$ synthetic emission maps, showing 
that their projected emission reflects the anisotropies of the blue pre-Wolf-Rayet wind. 
We demonstrate that projection effects play a non-negligible role in the shaping of 
observed Wolf-Rayet nebulae. According to the observer's viewing angle, these transcient 
objects can appear as a oblate form, a bipolar structure, an ellipse or a ring. 
%
%
Last, we discuss the potential origins of the pre-Wolf-Rayet circumstellar anisotropies. 
Two main scenarios have been identified so far, involving either a red or a blue 
supergiant star. However, we show that solely the blue supergiant one can, within the 
single star scenario, generate density distributions sufficiently asymmetric to channel 
the Wolf-Rayet stellar wind and to produce Wolf-Rayet nebulae's bipolarity. 
This lead us to question the well-accepted past evolution of the famous asymmetric 
bubble nebula NGC 6888 around the Wolf-Rayet WR 136, believed to be the archetypal 
resulting nebulae of a single red supergiant-to-Wolf-Rayet-evolving star. 
We propose that NGC 6888 might be driven by a multiple system, or underwent 
additional evolutionary phases. 
%


\section*{Acknowledgements}

\textcolor{black}{
The author thanks the referee A.~Raga from the University of Mexico for his numerous 
comments which greatly improved the quality of the paper. 
}
The author thanks L.~Oskinova from the University of Potsdam for discussions regarding 
evolved massive stars. 
The author acknowledges the North-German Supercomputing Alliance (HLRN) for providing 
HPC resources that have contributed to the research results reported in this paper.

\section*{Data availability}

This research made use of the {\sc pluto} code developed at the University of Torino  
by A.~Mignone and collaborators (\url{http://plutocode.ph.unito.it/}) and of the {\sc radmc-3d} 
code developed by C.~Dullemond and collaborators at the University of Heidelberg 
(\url{https://www.ita.uni-heidelberg.de/~dullemond/software/radmc-3d/}), respectively. 
The figures have been produced using the Matplotlib plotting library for the Python 
programming language (\url{https://matplotlib.org/}). 
The data underlying this article will be shared on reasonable request to the 
corresponding author.


\bibliographystyle{mn2e}

\footnotesize{
\bibliography{grid}
}


\end{document}